Modeling Compact Boron Clusters with the Next Generation of Environment-Dependent Semi-Empirical Hamiltonian

P. Tandy[†], Ming Yu, C. Leahy, C.S. Jayanthi, and S. Y. Wu*

Department of Physics and Astronomy, University of Louisville, Louisville, KY 40292, USA

*Email: sywu0001@louisville.edu

**Abstract.** A highly efficient semi-empirical Hamiltonian has been developed and applied to model the compact boron clusters with the intermediate size. The Hamiltonian, in addition to the inclusion of the environment-dependent interactions and electron-electron correlations with the on-site charge calculated self-consistently, has contained the environment-dependent excitation orbital energy to take into account the atomic aggregation effect on the atomic orbitals. The Hamiltonian for boron has successfully characterized the electron deficiency of boron and captured the complex chemical bonding in various boron allotropes including the planer and quasi-planer, the convex, the ring, the icosahedra, the fullerene-like clusters, the two-dimensional monolayer sheets, and the alpha boron bulk, demonstrating its transferability, robustness, reliability, and has the predict power. The Hamiltonian has been applied to explore the existence of the compact structure of boron clusters with the intermediate size. Over 230 compact clusters including the random, the rhombohedra, and the spherical icosahedra structures are obtained with the size up to 768 atoms. It has been found that, energetically, clusters containing most compacted icosahedra $B_{12}$ balls (i.e., the body-like rhombohedra clusters and trimmed spherical cut icosahedra clusters) are the most stable for large size ($N_{atom}$ >200) of boron clusters, while the spherical cut icosahedra, random structures, and cage-like boron clusters are competitive for the small or intermediate size ($24 < N_{atom} < 200$) of boron clusters.

†Current address: Defense Threat Reduction Agency, 8725 John J Kingman Rd, Stop 6201, Fort Belvoir VA, 22060-6201, USA

## 1. Introduction

Boron, analogue to its neighbor carbon element, has various allotropes with different bonding natures and structures in the crystalline and nanostructures. Especially, over the last several years the interest in boron nanostructures including the clusters, tubes, and monolayer sheets has grown dramatically either from the experimental synthesis or from the modeling through computer simulation studies. Most of the first-principles studies as well as the experimental synthesis have found that the small-sized boron clusters can have the form of the planer, the quasi-planer, and convex structures [1-11] and a transition to ring structures has been found [12-14]. The progress on discovery the stable small boron clusters has improved quite recently, for instant, the quasi-planer hexagonal $B_{36}^-$ and $B_{40}^-$ clusters have been observed [15, 16]. The theoretical postulated interesting boron clusters with the intermediate size, such as $B_{80}$ buckyball [17-23] and other cage-like clusters [24, 25] have also been postulated but not yet been observed. Therefore, searching stable inter-mediate size boron clusters with other type of structures, such as the compact clusters, became necessary from the computational molding using the more accurate first-principle calculations or the semi-empirical methods. Especially, it is desired to have the highly efficient and reliable semi-empirical methods for large amount simulations which is beyond the scope of the first-principle calculations for various configurations of the intermediate size of boron clusters.

The complicated nature of the 3 center bond, with deficiency of the un-occupied *p* orbitals [26], and the ability of boron to bond in so many different energetically competitive ways [27] (isomorphism), was too challenging for the traditional semi-empirical methods [28]. Some of the first attempts to model boron with the extended Hückel method involved modeling boron hydrides [29] and although that method has some limited accuracy, it is insufficient for any type of electronic structure calculation that would remotely compete with density functional theory (DFT). In this paper, we will develop an environment-dependent semi-empirical Hamiltonian (refereed as SCED-LCAO Hamiltonian) for boron element to characterize the electron deficiency of boron and capture the complex chemical bonding in various allotropes of boron. We will show in this paper that the SCED-LCAO Hamiltonian for boron is transferable, reliable, and robustness. It is not only able to model boron clusters, it can also model sheets, and extended structures accurately. We have applied this Hamiltonian to study various compact structures of boron clusters with the intermediate size.

The paper is organized as follows. We first introduce the next generation of the SCED-LCAO Hamiltonian in section 2. Next, we will discuss the transferability and the reliability of the Hamiltonian in section 3. The robustness as well as the applications to various allotropes of boron will be given in section 4. These results will demonstrate how well the SCED-LCAO Hamiltonian for boron characterizes the complex chemical bonding nature of boron in different environments. Then, we will present in section 5 our results on investigation on the stability, the energetics, and the structural properties of the intermediate size of compact boron clusters. A conclusion will be given in section 6.

## 2. Methodology

The original self-consistent and environment-dependent (SCED) Hamiltonian was constructed in the framework of linear combination of atomic orbitals (LCAO) [30-32]. It was designed to remedy the inadequacies of conventional semi-empirical Hamiltonians such as various versions of tight-binding Hamiltonians [28, 29]. Specifically, it allows the self-consistent determination of the charge re-distribution and includes environment-dependent multi-center interactions, two main deficiencies of conventional semi-empirical Hamiltonians that limit their applications to be system-specific and prevent those Hamiltonians to be transferrable. The inclusion of these new features makes the SCED-LCAO Hamiltonian transferrable, thus possessing predictive power. Furthermore parametric functions appearing in the SCED-LCAO Hamiltonian and the parameters there-in all have transparent physical meaning, allowing the physics of the results of calculations based on the SCED-LCAO Hamiltonian easily tractable.

We have constructed the SCED-LCAO Hamiltonians for Silicon, Carbon, Germanium, and Phosphorus and optimized the respective parameter sets using sets of judiciously chosen data bases [30-32]. We have also successfully tested our SCED-LCAO Hamiltonians for structural and electronic properties of Si-, C-, and Ge-based structures of different symmetries/geometries (bulk/3-d, surface/2-d, wires/1-d, clusters/0-d, *etc*) and different phases (diamond, fcc, bcc, *etc*) [30-38]. These tests have demonstrated the transferability and the reliability of the SCED-LCAO Hamiltonian in predicting properties of a wide range of Si-, Ge-, and C-based structures.

The diagonal and off-diagonal matrix elements of the SCED-LCAO Hamiltonian are given as follows:

$$H_{i\alpha,i\alpha}^{SCED-LCAO} = \varepsilon_{i\alpha} + (N_i - Z_i)U_i + \sum_{k\neq i}\left(N_k V_N(R_{ik}) - Z_k V_Z(R_{ik})\right) , \qquad (1)$$

$$H_{i\alpha,j\beta}^{SCED-LCAO} = \frac{1}{2}\{(\varepsilon_{i\alpha}^{'} + \varepsilon_{j\beta}^{'})K(R_{ij}) + [(N_i - Z_i)U_i + (N_j - Z_j)U_j]$$

$$+[\sum_{k\neq i}(N_k V_N(R_{ik}) - Z_k V(R_{ik})) + \sum_{k\neq j}(N_k V_N(R_{jk}) - Z_k V_Z(R_{jk}))]\}S_{i\alpha,j\beta} \quad (2)$$

where $\varepsilon_{i\alpha}$ can be identified as the orbital energy of the $\alpha$ -orbital corresponding to an isolated atom at the site *i*, $N_i$ the number of valence electrons associated with the atom at site *i*, $Z_i$ the number of valence electrons of the atom at the site *i*, $U_i$ the Hubbard-like energy representing the on-site correlation energy, $V_N(R_{ik})$ the electron-electron energy per number of electrons at site *k* between an electron distribution at site *i* and the electron distribution at site *k*, $V_Z(R_{ik})$ the electron-ion interaction energy between electrons associated with the atom at site *i* and the ion at site *k* per number of ionic charge, $K(R_{ij})$ a scaling function, and $S_{i\alpha,j\beta}$ the overlapping matrix. Specifically, we express $V_N(R_{ij})$ in terms of $V_Z(R_{ij})$ using a short range function $V_N(R_{ij}) = V_Z(R_{ij}) + \Delta V_N(R_{ij})$ . This is because both $V_N(R_{ij})$ and $V_Z(R_{ij})$ approach $e^2 / 4\pi\varepsilon_0 R_{ij}$ as $R_{ij} \to \infty$. The five parametric functions are given as follows:

$$K\left(R_{ij}\right) = e^{-\alpha_{ij,K} R_{ij}} \quad (3)$$

$$S_{ij,\tau}\left(R_{ij}\right) = \left(A_{ij,\tau} + B_{ij,\tau} R_{ij}\right) \frac{1 + e^{-\alpha_{ij,\tau} d_{ij,\tau}}}{1 + e^{-\alpha_{ij,\tau}\left(d_{ij,\tau} - R_{ij}\right)}} \quad (4)$$

$$V_{i,N}\left(R_{ij}\right) = V_{i,Z}\left(R_{ij}\right) + \Delta V_{i,N}\left(R_{ij}\right) \quad (5)$$

$$V_{i,Z}\left(R_{ij}\right) = \frac{E_0}{R_{ij}}\left(1 - \left(1 + B_{i,Z} R_{ij}\right) e^{-\alpha_{i,Z} R_{ij}}\right) \quad (6)$$

$$\Delta V_{i,N}\left(R_{ij}\right) = \left(A_{i,N} + B_{i,N} R_{ij}\right) \frac{1 + e^{-\alpha_{i,N} d_{i,N}}}{1 + e^{-\alpha_{i,N}\left(d_{i,N} - R_{ij}\right)}} \quad (7)$$

The parameters needed for the construction of the SCED-LCAO Hamiltonian include $\varepsilon'_{i\alpha}$, $U_i$ and the parameters characterizing the five parametric functions. $\varepsilon'_{i\alpha}$, instead of $\varepsilon_{i\alpha}$, appearing in the off-diagonal SCED-LCAO Hamiltonian matrix element, is a reflection of the difference in the environment of the offsite case vs. the onsite cases. These parameters are to be optimized with respect to an appropriately chosen data base highlighting properties of the system under consideration.

In practice, the orbitals involved in the SCED-LCAO Hamiltonian are limited to occupied atomic orbitals. Hence the SCED-LCAO scheme is a finite (incomplete) basis approach. Previous results indicate that the SCED-LCAO Hamiltonian works well for elements with sufficiently localized electrons, indicating that the presence of neighboring atoms in an atomic aggregate for these elements has not appreciably affected the occupation of atomic orbitals. However, to enable the SCED-LCAO Hamiltonian to have wider applicability, one has to address the issue related to the finite basis set. The usual approach is to add an excited orbital in the basis set as the interactions with neighboring atoms may excite the electron to unoccupied orbitals. However even with the addition of just a single excited orbital, it will lead to an unusually large increase in the number of parameters in the semi-empirical Hamiltonian, specifically in the overlapping matrix.

We decide to take a different approach. We take into consideration of possible occupation of the excited local atomic orbitals in an atomic aggregate by forming the *l*-like orbitals in the basis set to include the occupied "ground state" as well as the excited atomic orbitals corresponding to a certain quantum number *l* (with *l* standing for s, p, d, and etc.). It should be noted that this is still a semi-empirical approach with no explicit involvement of atomic orbitals other than the appearance of the quantum indices in the parametric functions representing the overlapping matrix. With this set up, the corresponding *l*-like orbital energy is expected to be different from the energy of the initially occupied *l* orbitals of the isolated atom. We model this shift in energy by the addition of a parametric function $W_{i\alpha}(R_{ij}) = W_{i\alpha}^0 e^{-\alpha_{i\alpha,W} R_{ij}}$ . It can be seen that $W_{i\alpha}(R_{ij})$ approaches zero for sufficiently large distance but at shorter distance the function is finite and alters the single atom result for the valence level term $\varepsilon_{i\alpha}$ accordingly. In this way, with the addition of a few more free parameters, the effects of the interactions with neighboring atoms can be included. Furthermore the incompleteness of the basis set can be improved by simply redefining *l*-like orbitals without explicitly adding any excited orbitals in the semi-empirical Hamiltonian.

The diagonal and off-diagonal matrix elements of the new and improved version of the SCED-LCAO Hamiltonian are then expressed as follows:

$$H_{i\alpha,i\alpha}^{SCED-LCAO} = \varepsilon_{i\alpha} + \sum_{k\neq i} W_{i\alpha}(R_{ik}) + (N_i - Z_i)U_i + \sum_{k\neq i}\left(N_k V_N(R_{ik}) - Z_k V_Z(R_{ik})\right) \ , \qquad (8)$$

$$H_{i\alpha,j\beta}^{SCED-LCAO} = \frac{1}{2}\{(\varepsilon_{i\alpha}^{'} + \varepsilon_{j\beta}^{'} + \sum_{k\neq i} W_{i\alpha}(R_{ik}) + \sum_{k\neq j} W_{j\alpha}(R_{jk}))K(R_{ij}) + [(N_i - Z_i)U_i + (N_j - Z_j)U_j]$$

$$+[\sum_{k\neq i}(N_k V_N(R_{ik}) - Z_k V(R_{ik})) + \sum_{k\neq j}(N_k V_N(R_{jk}) - Z_k V_Z(R_{jk}))]\}S_{i\alpha,j\beta} \qquad (9)$$

The total energy of the system is then given by

$$E_{total} = \sum_{\lambda}^{ooc} n_\lambda \varepsilon_\lambda + \frac{1}{2}\sum_i (Z_i^2 - N_i^2)U_i - \frac{1}{2}\sum_i \sum_{j\neq i} N_i N_j V_N(R_{ij}) + \frac{1}{2}\sum_i \sum_{j\neq i} Z_i Z_j \frac{E_0}{R_{ij}} \qquad (10)$$

To ensure the viability of the new and improved scheme and to test the transferability of the resulting Hamiltonian, we have subjected this scheme to a stringent test, the construction of the SCED-LCAO Hamiltonian for the element boron and the testing of its robustness and transferability. Boron compounds are known to form three-center two-electron (3c-2e) bonds where three atoms share two electrons. Semi-empirical Hamiltonians are not expected to perform well for boron-based systems, in particular their transferability [28, 29]. In this work, we have demonstrated the ability of the SCED-LCAO Hamiltonian to model boron and successfully tested its transferability. The test also suggests that the new version of the SCED-LCAO Hamiltonian should be applicable for elements with more delocalized electrons. In the following sections, we give in details the construction of the SCED-LCAO Hamiltonian for boron and the test for its transferability and robustness.

## 2. Transferability and reliability

Within the framework of the linear combination of $sp^3$ atomic orbitals, there are 25 parameters needed in the new version of the SCED-LCAO Hamiltonian for boron. They will be determined by minimizing the objective function, defined as the least-squares sum of the differences between the calculated properties and the reference values from the first-principle calculations, through a fitting process with an efficient global optimization algorithm (see the Appendix of Ref [30] for details). The optimized parameters of the SCED-LCAO Hamiltonian for boron are given in Table 1.

To ensure that the SCED-LCAO Hamiltonian for boron could characterize the complex chemical bonding nature of boron due to its electron-deficiency, we have built up a property database for the reference values in the fitting process including (1) 12 small boron clusters with planer, quasi-planer, convex, and ring structures, (2) the crystalline alpha boron phase with rhombohedra symmetry, (3) the monolayer alpha sheet (5 or 6 coordination number of boron atoms), and (4) the monolayer delta-4 sheet (4 coordination number of boron atoms) (see Ref [39] for the notations of the monolayer sheets). Especially, the isomer nature in some of the small clusters (e.g., $B_4$ with $D_{4h}$ and $D_{2h}$ symmetries, $B_6$ with $C_{5v}$ and $C_{2h}$ symmetries, $B_7$ with $C_{2v}$ and $C_{2h}$ symmetries, and $B_{12}$ with $C_{3v}$ and $D_{6h}$ symmetries, respectively) is carefully taken into account so that both the stable and the meta-stable structures can be characterized. The properties (i.e., the cohesive energy and the geometric properties) of these small boron clusters are listed in Table 2. The properties calculated by the SCED-LCAO Hamiltonian for boron using the optimized parameters are in good agreement with the results obtained from the *ab initio* calculations [40] (e.g., the CCSD (T) method with aug-cc-PVTZ basis [41]).

The transferability of the SCED-LCAO Hamiltonian for boron is ensured by adding the relative energy versus atomic volume curves for the extended systems, including the alpha boron [42, 43] and monolayer sheets [39, 44], into the properties database. The bonding nature in the alpha boron and monolayer boron sheets is different. The major chemical bond in the monolayer sheets is the type of the three-center. While, in the alpha boron, in additional to the three-center bonds, there is also the directional covalent two-center bonds between icosahedra $B_{12}$ [42, 43]. Fig. 1 shows the relative energy per atom (black curves denote SCED-LCAO results; dashed curves, DFT results) as a function of the ratio of atomic volume to the equilibrium atomic volume of alpha boron. Here the atomic volume for the monolayer sheets is defined as $V = Ah / N_{atom}$, where $A$ is the area of the unit cell of the sheet, $h$, the 'think' of the sheet (we assume it as the twice of the bond length), and $N_{\text{atom}}$, the number of atoms per unit cell, respectively. It can be seen clearly from Fig. 1 that the energy curves for the α boron and the monolayer α sheet using the SCED-LCAO Hamiltonian fit very well to those calculated from the first-principle method (i.e., the DFT-based VASP scheme with ultra soft pseudo-potential and GGA for electron correlations [45]). It is also worth noting that the shape and the minimum of the energy curve for the monolayer $\delta_4$ sheet is consistent to the DFT result except a slight shift of the relative energy by approximately 0.01 Hartree/atom.

## 3. Robustness and applications

To demonstrate that the SCED-LCAO Hamiltonian for the boron element is robustness, transferable, reliable, and has the predict power, we have applied the Hamiltonian with the optimized parameters to study several complex systems of boron. The molecular dynamics (MD) scheme based on the new version of the SCED-LCAO Hamiltonian has been employed in such calculations. The force on the *i*th atom are given by the following expression

$$F_i^l = -\sum_{\lambda}^{ooc}\sum_{i\alpha}\sum_{j\beta}(c_{i\alpha}^{\lambda})^* c_{j\beta}^{\lambda}\left[\frac{\partial H_{i\alpha,j\beta}}{\partial x^l} - \varepsilon_\lambda \frac{\partial S_{i\alpha,j\beta}}{\partial x^l}\right] + \frac{1}{2}\sum_i\sum_{j\neq i}\frac{\partial(N_i N_j V_N(R_{ij}))}{\partial x^l}$$
$$-\frac{1}{2}\sum_i\sum_{j\neq i} Z_i Z_j \frac{\partial(E_0/R_{ij})}{\partial x^l}, \quad l = x, y, z \quad . \tag{10}$$

The time step in the molecular dynamics simulations was set to be 1.2 fs and the force criteria for a fully relaxation process was set to be less than $10^{-2}$ eV/Å. The complex systems that we studied include (1) the energetics of $B_n$ ($n$ = 10-40), (2) the stability of the $B_{80}$ buckyball, and (3) the stability of monolayer triangular and alpha sheets.

### 4.1 The energetics of $B_n$ ($n$ = 10-40)

The trivalent boron has two *s* and one *p* valence electrons at the ground state. Its electron deficiency and the *s* and *p* orbitals hybridization allow it to form complex chemical bonds. The first–principle computational simulations and recent experimental observations [15, 16, 46-51] have shown that boron clusters within the size of 10-40 atoms can form various structures such as the quasi-planer, the convex, the icosahedra, and the ring shapes. Even for a given number of atoms, there are several isomers having quite different structures and energetics. There is no direct relation between the size of the boron clusters and the geometry shape of the cluster in this range of size. Therefore, to correctly characterize the chemical bonding nature, the energetics, and the structural properties of these boron clusters is a tremendous challenge for a semi-empirical Hamiltonian and a testimony for the reliability of such Hamiltonian. For this purpose, we have carried out MD simulations to study these boron clusters using the SCED-LCAO Hamiltonian for boron element developed above. The clusters concerned in this study include the quasi-planers of $B_{10}$, $B_{11,}$ $B_{15-2}$, $B_{19}$, and $B_{36-1;}$ the icosahedra of $B_{12}$; the convexes of $B_{13}$ isomers (e.g., $B_{13-1}$ and $B_{13-2}$), $B_{15-1}$, $B_{36-2}$, and $B_{40}$ isomers (e.g., $B_{40-1}$, $B_{40-2}$); and the rings of $B_{20}$ and $B_{24}$, respectively. It should note that all the stable quasi-planer and convex clusters are automatically obtained by directly relaxing the corresponding planer structures at 0 K. This demonstrates that, different from the DFT methods, such as the VASP scheme where some of the convex structures can not be directly obtained from the corresponding planer structures using relaxation process, the SCED-LCAO Hamiltonian is capable to drive a system from a metastable state (e.g., the planer) to a stable state (e.g., the quasi-planer) over a certain kinetic energy barrier. The obtained relative energies per atom with respect to that of $B_{40-2}$ are listed in the third column of Table 3. It is found that both the symmetry of the stabilized structures and the energy ordering among these clusters are in good consistent with the DFT-GGA results [45] (the fourth column of Table 3). The calculations in Table 3 are also consistent with other first-principle results (see Ref. [15, 16]). Especially, it is found that (1) the icosahedra $B_{12}$ is indeed unstable compared to its isomers (e.g., 0.107 eV/atom higher than the $B_{12}$ ring and 0.112 eV/atom higher than the $B_{12}$ quasi-planer) and other clusters with slight smaller size (e.g., $B_{10}$ and $B_{11}$), (2) the energy difference between two convexes of the isomers $B_{13-1}$ and $B_{13-2}$ are very small (~ $10^{-3}$ eV/atom) due to their similar structures. The SCED-LCAO Hamiltonian, however, can distinguish such tiny difference as well as the energy ordering. It is also true for other isomers such as the isomers of $B_{15-1}$ (convex) and $B_{15-2}$ (quasi-planer), the isomers of $B_{36-1}$ (quasi-planer) and $B_{36-2}$ (convex), and the isomers of $B_{40-1}$ (convex) and $B_{40-2}$ (convex), and (3) the $B_{36-2}$ is more stable than $B_{36-1}$, supporting the experimental report [15]. This testimony clearly demonstrates that the environment-dependence SCED-LCAO Hamiltonian for boron is capable of capturing the different structures of the boron clusters and is reliable.

**4.2 The stability of the $B_{80}$ buckyball**

Stimulated by the carbon $C_{60}$ fullerene, the $B_{80}$ backyball has been postulated from the *ab initio* calculations in 2007 [17]. This buckyball has the $C_{60}$ fullerene structure with 20 boron atoms at each center of the hexagons having the balance between the two-center and the three-center bonds. Since then, a large amount of interesting researches have focused on the stability of the $B_{80}$ buckyball [18-23]. The initial study reported that the stable $B_{80}$ has the $I_h$ symmetry, similar to that of $C_{60}$ fullerene [17]. After that, several reports have mentioned that the $B_{80}$ buckyball with $T_h$ symmetry (atoms at the center of the hexagons move inwards) is more stable than $I_h$ symmetry [18-22] and even the lower symmetry (e.g., $C_{2h}$ symmetry with atoms at the center of the hexagons move inwards/outwards alternately) is more stable than the $T_h$ symmetry [23]. Therefore, we took this challenge as a stringent robust test for our boron SCED-LCAO Hamiltonian. We run a SCED-LCAO-MD simulation for the $B_{80}$ buckyball starting from an $I_h$ structure. As shown in Fig. 2, $B_{80}$ with $I_h$ symmetry is indeed not stable. It quickly transforms to $T_h$ symmetry in about 0.3 ps. Furthermore, the backyball keeps $T_h$ symmetry in about 1.2 ps and finally stabilizes to $C_{2h}$ symmetry after 2.1 ps in total simulation time. The lowest energy state found with the current SCED-LCAO Hamiltonian was the $C_{2h}$ symmetry state with some atoms pointing out and others pointing in. The energy differences in these structures are exceedingly small, ~0.004eV/atom. It is clear seen that the boron SCED-LCAO Hamiltonian can provide accurate information about the stability versus the symmetry of the $B_{80}$ backyball and are consistent with the DFT calculations [17-23]. In fact, it directly predicts that the most stable $B_{80}$ backyball is with the $C_{2h}$ symmetry. Furthermore, we also carried out a thermal dynamics simulation on the $B_{80}$ buckyball by heating it up to 1800 K. We found that when the $B_{80}$ buckyball was slowly heated up to 1800 K (see Fig. 3), the $C_{2h}$ symmetry is totally distorted. This structures then was slowly cooled down to 0 K, and finally stabilized to a random structure with the cohesive energy about 0.052 eV/atom lower than $B_{80}$ backyball with the $C_{2h}$ symmetry, indicating that $B_{80}$ buckyball is a metastable structure and that is why it has not yet discovered experimentally. This result is consistent with the finding by the global geometry optimization on the density functional potential energy surface with the minim hopping algorithm [21].

### 4.3 The stability of monolayer triangular and alpha sheets

The existence and the stability of various boron monolayer sheets has been systematically studied using the first-principle particle-swam optimization global algorithm [39, 44]. It is found that the triangular sheet is stabilized to a buckled structure and the buckled α sheet is more stable than the flat alpha sheet. In fact, the buckled α sheet (referred as α'-sheet in Ref. [39]) is the most stable among these monolayer sheets [39]. To further validate the appropriateness of the SCED-LCAO Hamiltonian for boron, we have applied the SCED-LCAO based MD simulation to study the stabilities of the triangular and α sheets, as examples. As shown in Fig. 4, an initial flat triangular sheet of boron (the inset at the top left of Fig. 4) was allowed to fully relax (see the black curve in Fig. 4) during the MD simulation in a 2.4 ps. It clearly shows that the flat sheet is indeed unstable and becomes buckled after 0.36 ps. After 1.68 ps, the sheet is finally stabilized to a perfectly buckled triangular sheet with the buckling of 0.66 Å (see the inset at the middle right of in Fig.4). On the other hand, an initial flat α sheet (the inset at the bottom left of Fig. 4) stands for about 1.2 ps and then stabilizes to a slightly buckled structure (the inset at the bottom right of Fig. 4) in about 1.6 ps (see the red dashed curve in Fig. 4). This structure is about 0.01 (eV/atom) lower than the initial flat structure. These SCED-LCAO simulations reproduced the results from the DFT [39] and demonstrate again the reliability of the SCED-LCAO Hamiltonian.

The final robust check is relative energy ordering (refereed to the total energy per atom of the α boron) and the energy gaps for the boron two-/three-dimensional structures. They are the monolayer buckled triangular and α sheets, the flat $\delta_4$ and $B_{36}$ (also referred as borophene which was suggested in Ref [15]) sheets, and the α boron. The results are presented in the Table 4, together with the corresponding DFT results. It is found that energetically, the flat monolayer sheets (i.e., the flat $\delta_4$ and the $B_{36}$ sheets) are less stable than the buckled monolayer sheets (i.e., the buckled α and the triangular sheets). The α boron, on the other hand, is the most stable one among all the allotropes of boron. Furthermore, all the monolayer sheets are gapless materials, except the flat $B_{36}$ sheet (~ 0.13089 eV), but the α boron is semiconductor material (~ 1.9 eV). The highly consistency between the SCED-LCAO results and the DFT-GGA results calculated by VASP with ultra soft pseudo-potential [45] strongly demonstrate that the boron SCED-LCAO Hamiltonian is indeed robustness, transferable, reliable, and has the predict power. We will apply the Hamiltonian to model the intermediate size of boron clusters $B_n$ and predict the energetics of the boron clusters with the size of $n$ =100-800 atoms.

### 5. Energetics of intermediate sized compact boron clusters $B_N$

In the intermediate size, the cage, convex, and the fullerene-like structures have been theoretically predicted [17-25]. But, there are still questions for the intermediate sized boron clusters including (1) whether these cage-like structures are the most stable, especially at moderate temperature and (2) does there exists some other structures, such as the compact structure which exist and energetically stable? To answer these questions, we have applied the SCED-LCAO Hamiltonian for boron developed above to explore the existence of such interest compact boron clusters within the size up to 768 atoms. All of the compact structures concerned in this work are constructed by the distribution of the boron atoms in a topologically disordered pattern (referred as random), a rhombohedra cut from the alpha boron (referred as rhombohedra), a spherical cut from the alpha boron (referred as spherical), and removing surface atoms with more dangling bonds from the spherical cut clusters (referred as trimmed), respectively. These initial constructed clusters are first relaxed at 0 K with the power quenching scheme [52, 53], then heated to a moderate temperature (~ 1500 K) using simulated annealing, and finally slowly cooled down to 0 K. The obtained stable compact boron clusters, their structural, and relative energies will be discussed in the followings.

### 5.1 Random structures of $B_N$ (Random)

We have obtained over 90 stable boron clusters $B_N$ ($N$ =80~ 228) with random structures. The stable random structures clusters of $B_{101}$ (Random), $B_{228}$ (Random), and $B_{230}$ (Random) are shown in Figure 5, as an example. The total energy per atom of the boron clusters as a function of the number of atoms has been plotted in the Figure 6 (see the pink open circles in Fig. 6). The energy varies with the size of the clusters with a fluctuation of ~0.07 eV/atom. The total energy per atom of the $B_{80}$ buckyball is also shown in Fig.6 with the black star for comparison. It is clearly seen that most amorphous clusters around that size are fairly lower in energy than $B_{80}$ buckyball [22, 51, 54-56] (e.g., the total energy of $B_{80}$ (Random) is -39.70 eV per atom and that of $B_{80}$ buckyball is -39.6276 eV per atom, respectively), indicating that for the intermediate size of boron clusters, the cage-like structures are less stable than the compact structures. This does not mean that the $B_{80}$ buckyball cannot exist. But, it does imply that any laboratory fabrication technique, for boron cluster construction, that involves significant heat; the symmetric structures that could be produced will have to compete with boron's natural tendency to enjoy being in an amorphous state.

### 5.2 Rhombohedra structures of $B_N$ (i_j_k)

Each rhombohedra structure of compact boron cluster is constructed by a rhombohedra cut from the alpha boron. It is labeled with $B_N$ (i_j_k) where the indices designate how many icosahedra $B_{12}$ ball there are in the cluster and the total number of atoms in the cluster is given by 12 x i x j x k. Apparently, a sheet-like rhombohedra cluster has the form of $B_N$ (i_j_1) with various values of *i* and *j*, and a chain-like rhombohedra cluster has the form of $B_N$ (i_1_1) with *i* icosahedra balls along a given direction. Over 80 of such rhombohedra clusters $B_N$ (i_j_k) with *N* range from 24 to 768 are constructed and fully relaxed. It is found that the stabilized compact clusters still keep the basic rhombohedra structures which, for example, can be seen from the insets of Fig. 6 such as the chain-like $B_{108}$ (9_1_1) (see the inset at the top of Figure 6), the sheet-like $B_{336}$ (4_7_1) (see the inset in middle of Figure 6), and the bulk-like $B_{768}$ (4_4_4) (see the inset at the bottom of Fig. 6). This is clear because the icosahedra ball is the building block of the alpha boron. Its bonding nature including the three-center bond inside the icosahedra ball and the two-center bond between icosahedra balls in the rhombohedra symmetry makes the alpha boron the most stable structures among the boron allotropes. The compact clusters with the rhombohedra cut from the alpha boron keep the same symmetry as the bulk phase. Therefore, they prefer to stay with such symmetry with only minor distortion on the surface.

Energetically, it is found that the chain-like $B_N$ (i_1_1) (see the black curve with open circles in Fig. 6) and sheet-like $B_N$ (i_j_1) (see the red and green curves with open squares and diamonds in Fig. 6) rhombohedra clusters are less stable than the clusters with random structures within the size of $N$<200. The bulk-like rhombohedra $B_N$ (i_j_k) (see the red, green, and black curves with solid squares, diamonds, and circles in Fig. 6) clusters, however, have similar energy for $N$<100 and much lower energy for $N$>100 than that of $B_N$ (Random) cluster. For $N$>300, the some of the sheet-like rhombohedra clusters (e.g., $B_N$ (i_7_1)) become much stable. Unlike the $B_N$ (Random) cluster with a fluctuated energy, the energy of the rhombohedra clusters decreases monotonically with the increase of the number of atoms. The energy ordering and the energy trend to the infinite size of the rhombohedra clusters strongly depend on the shape of the icosahedra cluster. In the case of the chain-like rhombohedra clusters $B_N$ (i_1_1), the energy decreases quickly for small $N$ and approaches to a constant after $N$ > 100 which is expect to that of the infinite icosahedra chain structure. In the case of the sheet-like rhombohedra clusters $B_N$ (i_j_1), the energy gradually decreases as increasing the size of the cluster. But the energy of $B_N$ (i_2_1) (the red curve with the open red-squares) and $B_N$ (i_7_1) (the green curve with the open green-diamonds) will approach to different energy values at their infinite limits, one is to the value of a infinite ribbon with the width of two icosahedra balls, and the other is to the value with the width of seven, respectively. The wider the sheet-like icosahedra cluster is, the lower its the energy is. This makes sense since the wider sheet-like rhombohedra cluster will has less edge effect, and is more stable. Similarly, in the case of the bulk-like icosahedra clusters $B_N$ (i_j_k), for a given number of atoms, the $B_N$ (i_3_2) (the green curve with solid green-diamonds) is energetically lower than $B_N$ (i_2_2) (the red curve with solid red-squares). But the $B_N$ (i_i_i) (the black curve with solid black-circles) is the lowest in energy and its energy will definitely approach to the value of the alpha boron (-40.12eV/atom) when the number of the atom $N$ goes to the infinite.

### 5.3 Spherical icosahedra structures of $B_N$

In additional to the rhombohedra cut from the alpha boron, another type of interesting and possible existing compact boron clusters are those constructed by the spherical cut from the alpha boron. There are two possible ways to perform the spherical cut: one is centered on an icosahedra ball (referred as $B_N$ (Ball)), and the other is centered between icosahedra balls (referred as $B_N$ (Empty)). We have constructed 59 spherical icosahedra clusters and performed MD simulations to relax the strains generated from the initial construction. Some of the stabilized clusters are shown in Fig. 7. For the spherical cut clusters, since the interior are formed by the icosahedra balls and the surface are opened icosahedra balls, the surface atoms are not stable and will reconstruct during the relaxation. It can be seen that each cluster centered upon icosahedra ($B_N$ (Ball) in Fig. 7) has at least one icosahedra ball in the interior. As the size of the cluster increases, the number of icosahedra balls in the interior of the cluster increases. The distortion of those interior balls depends on the size of the cluster. For the small size, such as $B_{30}$ (Ball) and $B_{54}$ (Ball), there is one icosahedra ball at the center, and the surface atoms reconstructed to a cage-like structure. This type of clusters is analogy to the boron cage with an icosahedra inserted as investigated by other theoretical groups [57, 58]. They have shown that such type of clusters is more stable than the cage-like boron clusters [57, 58]. As for the increase of the size of $B_N$ (Ball) clusters, there are more icosahedra balls in the interior and the distortion of the interior icosahedra will be less and less. The surface reconstructions will also possess a certain symmetric form, as can been seen from the left panel in Fig. 7. While in the case of the $B_N$ (Empty) (see the right panel in Fig. 7), since there is a hole at the center of such clusters, the atoms in the interior need to bond each other. This relaxation procedure together with the dangling bonds of surface atoms drives the cluster to a more distorted and compact structure with low symmetry.

Energetically, these spherical cut icosahedra clusters (see the blue open up-triangles for $B_N$ (Ball) and the orange open down-triangles for $B_N$ (Empty) in Fig. 6) have lower energy than the chain-like rhombohedra $B_N$ (i_1_1) clusters and even the sheet-like rhombohedra $B_N$ (i_j_1) clusters. That means, for the intermediate size of boron clusters, the 3D-like compact structures (e.g., bulk-like rhombohedra $B_N$ (i_j_k)) and the spherical icosahedra $B_N$ (Ball/Empty) are more stable than the less bulk-like compact structures (e.g., the chain- and sheet-like rhombohedra $B_N$ (i_1_1) and $B_N$ (i_j_1) clusters). Furthermore, for a size of the spherical cut cluster below 100 atoms, the $B_N$ (Empty) clusters are lower in energy than the $B_N$ (Ball) clusters. In the range of 100-300 atoms the $B_N$ (Ball) clusters, however, become lower in energy. They are even lower than random structures of boron clusters $B_N$ (Random) (pink open circles in Fig. 6) in energy. This result again demonstrates that for the intermediate size of the boron clusters (i.e., $N$>200) the more body-like compact structure is more stable than the random structures and the cage-like structures. As the size of the cluster over 400 atoms, both types of the spherical icosahedra clusters tend to have a similar in energy. This makes sense because when the size of the cluster is larger than 400 atoms both types of the spherical cut icosahedra clusters have lots of complete icosahedra ball in the interior so they are competitive in energy.

In the alpha boron clusters with spherical cut, surface atoms left over from the spherical cut form incomplete or open icosahedra balls. These surface atoms have many dangling bonds and then tend to bond each other during the relaxation. If instead those atoms are removed from the cut cluster, we would have a compact shape of icosahedra with no partial icosahedral structures (referred as $B_N$ (Trimmed)). An example of such trimmed cluster is shown in Fig. 8. It is $B_{288}$ (Trimmed) with 19 icosahedra balls in it. It does not look spherical because its size is too small (the top view of this structure shows hexagon symmetry, and the side view is a square-like shape), but it is the most compact cluster tested in this work cut from the alpha boron. There are no open icosahedra balls on the cluster surface and each surface icosahedra ball has at least three inter-icosahedra bonds to the neighbors. As expected, after the relaxation, the outer corners do reconstruct slightly, but the general shape stays the same. It is found that this cluster is the lowest in energy (see the black open square at $N$ =228 in Fig. 6) and therefore the most stable cluster among what we have ever found for an alpha boron cut among the similar size. It is compact, fully made of icosahedra, spherical and resistant to reconstruction up to about 1000K.

**5.4 Structural properties of boron clusters**

We have obtained abut 230 compact boron clusters with the random structures, the rhombohedra structures, the spherical icosahedra structures, and the trimmed spherical icosahedra structures, respectively. The energetics of these compact boron clusters, as shown in Fig. 6, strongly depend their structural properties. To shed light into the relationship between their relative energy and structural property, we performed local structural analysis and compared several different structures of similar size boron clusters to the alpha boron as the benchmark. Six different types of such clusters around ~220 atoms have been selected. They are the sheet-like rhombohedra cluster $B_{216}$ (9_2_1), the bulk-like rhombohedra cluster $B_{216}$ (3_3_2), the random structures cluster $B_{228}$ (Random), the spherical icosahedra cluster $B_{216}$ (Ball) and $B_{210}$ (Empty), and the trimmed spherical icosahedra cluster $B_{228}$ (Trimmed), respectively. The average bond length, the average neighbors, and the relative energy per atom to the alpha boron are listed for each of these clusters in Table 5. We found that the trimmed $B_{228}$ is energetically favorable among these size similar clusters. The sheet-like $B_{216}$ (9_2_1), on the other hand, is energetically unfavorable. This result indicates that the more compact clusters with higher symmetry and less dangling bonds on the surface atoms are more stable. Compared with the alpha boron (the last row of the Table 5), it can be seen that, the average bond length of boron clusters is shortened, the average neighbors of boron atoms is increased (except the case of the sheet-like rhombohedra $B_{216}$ (9_2_1) where many boron atoms are at the edges). As we know the alpha boron is built by the icosahedra ball and has rhombohedra symmetry. The bond length between boron atoms inside the icosahedra ball (referred as intra-bonds) is either 1.74Å or 1.80 Å, and the bond length between boron atoms in the neighboring icosahedra balls (referred as inter-bonds) is 1.67 Å. There are six inter-bonds per icosahedra $B_{12}$ ball. Thus, six boron atoms of an icosahedra ball are bonded to the boron atoms in the six neighboring icosahedra balls. While, the remaining other six boron atoms in the icosahedra ball have interactions with another six neighboring icosahedra balls with the distance of 2.02 Å (referred as the next-nearest-neighbor (NNN) distance). The average neighbor of a boron atom, therefore, is 5.50 within the cut-off of 2.0 Å (shown in Table 5). What we found is that when we truncated the alpha boron (either rhombohedra cut or spherical cut) to form the compact boron clusters (1) the intra-bond lengths are distributed in a wider range (e.g., 1.62-1.83 Å in $B_{216}$ (9_2_1); 1.52-1.73 Å in $B_{210}$ (Empty); 1.57-1.81 Å in the $B_{216}$ (Ball); 1.64-1.77 Å in the $B_{228}$ (3_3_2);1.58-1.80 Å in the $B_{228}$ (Trimmed)), (2) the inter-bond lengths are slightly extended (e.g., 1.71-1.76 Å in $B_{216}$ (9_2_1), 1.63-1.76 Å in $B_{210}$ (Empty); 1.69-1.78 Å in the $B_{216}$ (Ball); 1.70-1.89 Å in the $B_{228}$ (3_3_2);1.69-1.87 Å in the $B_{228}$ (Trimmed)), and (3) the number of neighbors is increased because the distances of NNN atoms are shortened and even bonded in the compact clusters as the result of the surface reconstruction. In particular, as can be seen from the third column of Table 5, the spherical cut centered at Ball and the trimmed clusters are much compacted (the average neighbors is 5.81 in $B_{216}$ (Ball) and 5.842 in $B_{228}$ (Trimmed), respectively) than the rhombohedra cut (the average neighbors is 5.379 in $B_{216}$ (9_2_1) and 5.676 in $B_{216}$ (3_3_2), respectively) and spherical cut centered at the Empty (the average neighbors is 5.57 in $B_{210}$ (Empty)). The $B_{228}$ (Random) cluster, on the other hand, is quite different from the clusters cut from the alpha boron. It has random (or amorphous) structure and a more compact with average neighbors of 5.833.

By comparing the pair distribution functions of these clusters to the corresponding alpha boron we found that the first peak in the pair-distribution function of the alpha boron (black dotted curve in Figure 9) is broadened in the clusters; in particular, the shoulder of the first peak around 2.02 Å in the alpha boron (representing the NNN distance between the neighboring icosahedra balls) is smoothed out in the clusters. This is, as we discussed above, the shortening of the bond between NNN atoms in the compact clusters due to the structure distortion and the surface reconstruction. The sharp peaks nature at larger distance in alpha boron is the tell-tale signature of crystalline alpha boron; this is not present in clusters, instead we see a broadened single peak in all of the clusters examined; in particular, the most broaden peak around 3.0 Å is found in the $B_{228}$ (Random), indicating its randomness feature.

A similar phenomenon is found in the angular distribution function (Fig. 10). Boron alpha is comprised of many $60^0$ degree angles due to the numerous equilateral triangles in the icosahedra. Secondly, the triangles do not lay flat. They are wrapped over a ball, the angle they make as they bend is $109^0$ degrees, the angle that is made between the icosahedra and connection line to the next icosahedra is about $90^0$ and $115^0$ degrees. This accounts for the 3 major angles in boron alpha. We also see the same $60^0$ degree bonding heavily in all of the boron clusters. This makes perfect sense, as we see the true nature of boron with heavy three center triangular bonding. But by comparing the peak at $60^0$ degree for the alpha boron to those for boron clusters, we found that this peak is broadened in the clusters indicating the large distortion of the icosahedra ball in the clusters. The broadest peaks were found in $B_{228}$ (random) which clearly demonstrate the amorphous structure. In conclusion it is noted that the lowest energy cluster observed in the large size of the boron clusters was that of a compact spherically shaped boron cluster comprised solely of 12 atom icosahedra. All pure boron clusters exhibit similar bonding as evidenced by the local analysis.

## 6. Conclusion

We have demonstrated the SCED-LCAO Hamiltonian for boron is transferable, reliable, and has the predictive power. It successfully characterizes the electron deficiency of boron and captures its complex chemical bonding in various allotropes of boron including the planer and quasi-planer, the convex, the ring, the icosahedra, and the fullerene-like clusters, the two-dimensional monolayer sheets with triangular, alpha, delta-4, and $B_{36}$ based symmetries, and the alpha boron bulk. In particular, it is capable to directly drive, using the MD version of the SCED-LCAO method, (1) the $B_{80}$ buckyball from high symmetry ($I_h$) with high energy to the low symmetry ($C_{2h}$) with low energy, (2) from the $C_{2h}$ symmetry to a random through a thermodynamics procedure structures, and (3) the flat triangular/alpha sheet to more stable buckled sheet. All the results including the energetics, the bonding, and the geometry of the boron allotropes studied are in good consistent with the DFT calculations as well as the experimental observations. The boron Hamiltonian was applied to model the intermediate size compact boron clusters $B_N$ ($N$ = 100 – 768) with random, rhombohedra, spherical icosahedra, and trimmed spherical icosahedra structures. It is clear to find that (1) the chain-like rhombohedra clusters $B_N$ (i_1_1) are energetically higher than any other type of compact boron clusters, (2) the sheet-like rhombohedra clusters $B_N$ (i_j_1) are lower than the chain-like clusters in energy, (3) while the spherical cut alpha boron clusters and the random structure clusters are even energetically lower than the sheet-like clusters, (4) but the lowest in energy among all types of the boron clusters are $B_N$ (i_j_k) bulk-like icosahedra and trimmed spherical icosahedra clusters. Furthermore, the cage-like structures, e.g., the $B_{80}$ buckyball, is competitive in energy only with random structures $B_N$ (Random). These results indicate that the compact boron clusters which contain most compacted icosahedra balls (i.e., the bulk-like rhombohedra $B_N$ (i_j_k) clusters and trimmed spherical icosahedra clusters) are the most stable for large size ($N_{atom}$ >200) of boron clusters, while the spherical cut, random structures, and cage-like boron clusters are competitive for the small or intermediate size ($24 < N_{atom} < 200$) of boron clusters.

**Acknowledgement:** The first author thanks for the support by the Dillon Fellowship at University of Louisville. We also acknowledge Cherno B Kah for his assistance in preparing the manuscript.

References

[1] M. Atiş, C. Özdoğan, and Z.B. Güvenç, International Journal of Quantum Chemistry **107**, 729 (2007)

[2] Ihsan Boustain, Zhen Zhu, and David Tománek, Phys. Rev. B **83**, 193405 (2011)

[3] Constantin Romanescu, Dan J. Harding, André Fielicke, and Lai-Sheng Wang, J. Chem. Phys. **137**, 014317 (2012)

[4] I. Boustani, Chemical Physics Letters **233**, 273 (1995)

[5] I. Boustani, Surf. Sci. **370**, 355 (1997)

[6] I. Boustani, J. Solid State Chemistry **133**, 182 (1997)

[7] [47'] H. J. Zhai, B. Kiran, J. Li, L. S. Wang, Nat. Mater. **2**, 827 (2003)

[8] H. J. Zhai, A. N. Alexandrova, K. A. Birch, I. A. Boldyrev, and L. S. Wang, Angew. Chem., Int. Ed. **42**, 6004 (2003)

[9] W. Huang, A. P. Sergeeva, J. H. Zhai, B. B. Averliev, L. S. Wang, Nat. Chem. **2**, 202 (2010)

[10] A. N. alexandrova, A. I. Boldyrev, H. J. Zhai, and L. S. Wang, Coord. Chem. Rev. **250**, 2811 (2006)

[11] R. T. Galeev, Q. Chen, J. G. Guo, H. Bai, C. Q. Miao, H. G. Lu, P. A. Sergeeva, S. D. Li, and A. I. Boldyrev, Phys. Chem. Chem. Phys. **13**, 11575 (2011)

[12] B. Kiran, S. Bulusu, H. J. Zhai, S. Yoo, X. C. Zeng, and L. S. Wang, Proc. Natl. Acd. Sci. U.S.A. **4**, 916 (2005)

[13] E. Oger, N. R. M. Crawford, P. Kelting, P. weis, M. M. Kappes, R. Ahlrichs, Angew. Chem., Int. Ed. **46**, 8503 (2007)

[14] W. An, S. Bulusu, Y. Gao, X. C. Zeng, J. Chem. Phys. **124**, 15430 (2006)

[15] Zachary A. Piazza, Han-Shi Hi, Wei-Li Li, Ya-Fan Zhao, Jun Li, and Lai-Sheng Wang, Nature Communication, DOI: 20.1038/ncomms4113 (2014)

[16] Hua-Jin Zhai, Ya-Fan Zhao, Wei-Li Li, Qian Chen, Hui Bai, Han-Shi Hu, Zachary A. Piazza, Wen-Juan Tian, Hai-Gang Lu, Yan-Bo Wu, Yue-Wen Mu, Guang-Feng Wei, Zhi-Pan Liu, Jun Li, Si-Dian Li, and Lai-Sheng Wang, Nat. Chem. DOI: 10.1038/NCHM.1999 (2014)

[17] N. Gonzalez Szwacki, A. Sadrzadeh, and B.I. Yakobson, Phys. Rev. Lett. **98**, 166804 (2007)

[18] G. Gopakumar, M.T. Nguyen, and A. Ceulemans, Chemical Physics Letters **450**, 175 (2008)

[19] A. Ceulemans, J. Tshishimbi, G. Gopakumar, and M. T. Nguyen, Chemical Physics Letters **461**, 226 (2008)

[20] Tunna Baruah, Mark R. Pederson, and Rajendra R. Zope, Phys. Rev. B **78**, 045408 (2008)

[21] S. De, A. Willand, M. Amsler, P. Pochet, L. Genovese, and S. Goedecker, Phys. Rev. Lett. **106**, 225502 (2011)

[22] Rosi N. Gunasinghe, Cherno B. Kah, Kregg D. Quarles, and Xiao-Qian Wang, Appl. Phys. Lett. **9**, 261906 (2011)

[23] N. Gonzalez Szwacki, A. Sadrzadeh, and B.I. Yakobson, Phys. Rev. Lett. **100**, 159901 (2008)

[24] Nevill Gonzalez Szwacki, Nanoscale Res. Lett. **3**, 49 (2008)

[25] Rajendra R. Zole, Europhysics Letters **85**, 68005 (2009)

[26] I. Boustani, Chem. Phys. Lett. **240**, 135 (1995)

[27] H. Kato, K. Yamashita, and K. Morokuma, Chem. Phys. Lett. **190**, 361 (1992)

[28] A.D. Lilak, S.K. Earles, K.S. Jones, M.E. Law, and M.D. Giles, *Electron Devices Meeting, 1997. IEDM ’97. Technical Digest., International*: IEEE, pp. 493–496 (1997)

[29] R. Hoffmann and W.N. Lipscomb, J. Chem. Phys. **37**, 2872 (1962)

[30] C. Leahy, M. Yu, C.S. Jayanthi, and S.Y. Wu, Phys. Rev. B **74**, 155408 (2006)

[31] S.-Y. Wu, C.S. Jayanthi, M. Yu, and C. Leahy, *Handbook of Materials Modeling*, edited by S.Yip, Vol. I, Springer series, p2929, Netherland (2005)

[32] M. Yu, S.Y. Wu, and C.S. Jayanthi, Physica E **42**, 1 (2009)

[33] Ming Yu, Indira Chaudhuri, C. Leahy, C.S. Jayanthi, and S.Y. Wu, J. Chem. Phys. **130**, 184708 (2009)

[34] W.Q. Tian, M. Yu, C. Leahy, C.S. Jayanthi, and S.-Y. Wu, Journal of Computational and Theoretical Nanoscience **6**, 390 (2009)

[35] M. Yu, C.S. Jayanthi, and S.Y. Wu, Nanotechnology **23**, 235705 (2012)

[36] M. Yu, C. S. Jayanthi, and S. Y. Wu, J. Material Research **28**, 57 (2013)

[37] Z. H. Xin, C. Y. Zhang, M. Yu, C. S. Jayanthi, and S. Y. Wu, Computational Materials Science **84**, 49 (2014)

[38] I. Chaudhuri **,** Ming Yu, C. S. Jayanthi, and S. Y. Wu, J. Phys. Condensed Matter **26**, 115301 (2014)

[39] Xiaojun Wu, Jun Dai, Yu Zhao, Zhiwen Zhou, Jinlong Yang, and Xiao Cheng Zeng, ACS Nano **6**, 7443 (2012)

[40] Gaussian 03, Revision C.02, M. J. Frisch, G. W. Trucks, H. B. Schlegel, G. E. Scuseria, M. A. Robb, J. R. Cheeseman, J. A. Montgomery, Jr., T. Vreven, K. N. Kudin, J. C. Burant, J. M. Millam, S. S. Iyengar, J. Tomasi, V. Barone, B. Mennucci, M. Cossi, G. Scalmani, N. Rega, G. A. Petersson, H. Nakatsuji, M. Hada, M. Ehara, K. Toyota, R. Fukuda, J. Hasegawa, M. Ishida, T. Nakajima, Y. Honda, O. Kitao, H. Nakai, M. Klene, X. Li, J. E. Knox, H. P. Hratchian, J. B. Cross, V. Bakken, C. Adamo, J. Jaramillo, R. Gomperts, R. E. Stratmann, O. Yazyev, A. J.

Austin, R. Cammi, C. Pomelli, J. W. Ochterski, P. Y. Ayala, K. Morokuma, G. A. Voth, P. Salvador, J. J. Dannenberg, V. G. Zakrzewski, S. Dapprich, A. D. Daniels, M. C. Strain, O. Farkas, D. K. Malick, A. D. Rabuck, K. Raghavachari, J. B. Foresman, J. V. Ortiz, Q. Cui, A. G. Baboul, S. Clifford, J. Cioslowski, B. B. Stefanov, G. Liu, A. Liashenko, P. Piskorz, I. Komaromi, R. L. Martin, D. J. Fox, T. Keith, M. A. Al-Laham, C. Y. Peng, A. Nanayakkara, M. Challacombe, P. M. W. Gill, B. Johnson, W. Chen, M. W. Wong, C. Gonzalez, and J. A. Pople, Gaussian, Inc., Wallingford CT, 2004.

[41] P.A. Denis, Chemical Physics Letters **395**, 12 (2004)

[42] R.J. Nelmes, J.S. Loveday, D.R. Allan, J.M. Besson, G. Hamel, P. Grima, and S. Hull, Phys. Rev. B **47**, 7668 (1993)

[43] J. Yang, Y. Yang, S.W. Waltermire, X. Wu, H. Zhang, T. Gutu, Y. Jiang, Y. Chen, A.A. Zinn, R. Prasher, T.T. Xu, and D. Li, Nat Nano **7**, 91 (2012)

[44] H. Tang and S. Ismail-Beigi, Phys. Rev. B **82**, 115412 (2010)

[45] G. Kresse and J. Hafner, Phys. Rev. B **48**, 13115 (1993)

[46] T.B. Tai, D.J. Grant, M.T. Nguyen, and D.A. Dixon, Journal of Phys. Chem. A **114**, 994 (2009)

[47] H.T. Pham, L.V. Duong, B.Q. Pham, and M.T. Nguyen, Chemical Physics Letters **577**, 32 (2013)

[48] D.Y. Zubarev and A.I. Boldyrev, Journal of Computational Chemistry **28**, 251 (2007)

[49] S. Mukhopadhyay, H. He, R. Pandey, Y.K. Yap, and I. Boustani, Journal of Physics: Conference Series, p. 012028 (2009)

[50] B. Kiran, S. Bulusu, H.J. Zhai, S. Yoo, X.C. Zeng, and L.S. Wang, Proceedings of the National Academy of Sciences of the United States of America **102**, 961 (2005)

[51] N.G. Szwacki, Nanoscale Research Letters **3**, 49 (2007)

[52] D.A. Drabold, P.A. Fedders, S. Klemm, and O.F. Sankey, Phys. Rev. Lett. **67**, 2179 (1991)

[53] P.A. Fedders, Phys. Rev. B **52**, 1729 (1995)

[54] R.R. Zope, T. Baruah, K.C. Lau, A.Y. Liu, M.R. Pederson, and B.I. Dunlap, Phys. Rev. B **79**, 161403 (2009)

[55] J.T. Muya, F. De Proft, P. Geerlings, M.T. Nguyen, and A. Ceulemans, J. Phys. Chem. A **115**, 9069 (2011)

[56] R.G. Delaplane, U. Dahlborg, W.S. Howells, and T. Lundström, Journal of Non-Crystalline Solids **106**, 66 (1988)

[57] Jijun Zhao, Lu Wang, Fengyu Li, and Zhougfang Chen, J. Phys. Chem. A **114**, 9969 (2010)

[58] Dasari L. V. K. Prasad and Eluvathingal D. Jemmis, Phys. Rev. Lett. **100**, 165504 (2008)

Table1 The optimized SCED-LCAO Hamiltonian parameters for boron

| Symbols | Values | Symbols | Values |
|---|---|---|---|
| $\varepsilon_s$ | -13.460 (eV) | $d_N$ | -0.597 (Å) |
| $\varepsilon_p$ | -8.430 (eV) | $B_{ss\sigma}$ | 0.318 (Å$^{-1}$) |
| $\varepsilon_s^{'}$ | -16.411 (eV) | $\alpha_{ss\sigma}$ | 1.477 (Å$^{-1}$) |
| $\varepsilon_p^{'}$ | -14.529 (eV) | $d_{ss\sigma}$ | 0.520 (Å) |
| $W_s^0$ | -0.921 (eV) | $B_{sp\sigma}$ | 0.466 (Å$^{-1}$) |
| $W_p^0$ | 0.183 (eV) | $\alpha_{sp\sigma}$ | 1.819 (Å$^{-1}$) |
| $\alpha_{s,W}$ | 2.172 (Å$^{-1}$) | $d_{sp\sigma}$ | 1.118 (Å) |
| $\alpha_{p,W}$ | 1.225 (Å$^{-1}$) | $B_{pp\sigma}$ | -0.906 (Å$^{-1}$) |
| $\alpha_K$ | 0.173 (Å$^{-1}$) | $\alpha_{pp\sigma}$ | 3.634 (Å$^{-1}$) |
| $U$ | 18.586 (eV) | $d_{pp\sigma}$ | 1.529 (Å) |
| $B_Z$ | 2.917 (Å$^{-1}$) | $B_{pp\pi}$ | -0.305 (Å$^{-1}$) |
| $A_N$ | -2.075 (eV) | $\alpha_{pp\pi}$ | 1.425 (Å$^{-1}$) |
| $B_N$ | -1.143 (Å$^{-1}$) | $d_{pp\pi}$ | 0.326 (Å) |
| $\alpha_N$ | 2.502 (Å$^{-1}$) | $R_{cut}$ | 7.0 (Å) |

Table 2 Comparisons of cluster properties (geometry and cohesive energies) of $B_N$ ($N \leq 12$) calculated from the SCED-LCAO method and the CCSD(T)/aug-cc-PVTZ method [41] as implemented in the Gaussian package [40]. Note that for each cluster coordinates of only in-equivalent atoms are given.

| $B_N$ | Symmetry | Structure | SCED-LCAO values | *ab initio* values[16, 24] |
|---|---|---|---|---|
| $B_2$ | $D_{ih}$ | 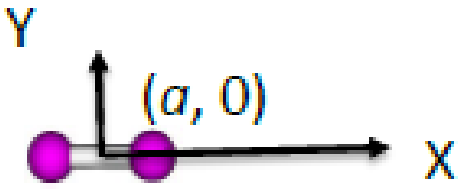 | $a$ = 0.882 Å<br>$E$ = -1.082 eV | $a$ = 0.818 Å<br>$E$ = -1.019 eV |
| $B_3$ | $D_{3h}$ | 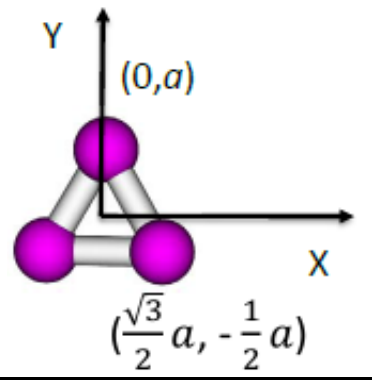 | $a$ = 0.857 Å<br>$E$ = -2.434 eV | $a$ =0.890 Å<br>$E$ = -2.747 eV |
| $B_4$ | $D_{4h}$ | 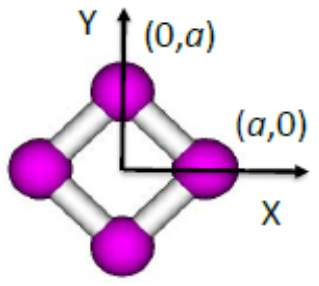 | $a$ =1.039 Å<br>$E$ = -2.968 eV | $a$ =1.070 Å<br>$E$ = -3.352 eV |
| $B_4$ | $D_{2h}$ | 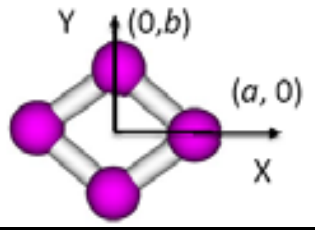 | $a$ =1.066 Å<br>$b$ =1.013 Å<br>$E$ = -2.969 eV | $a$ =1.193 Å<br>$b$=0.939 Å<br>$E$ = -3.361 eV |
| $B_5$ | $C_{2v}$ | 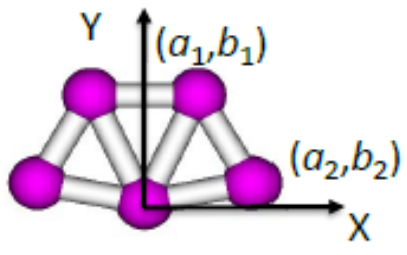 | $a_1$ =0.266 Å<br>$a_2$ =1.578 Å<br>$b_1$=1.547 Å<br>$b_2$=0.762 Å<br>$E$ = -3.098 eV | $a_1$ =0.302 Å<br>$a_2$ =1.655 Å<br>$b_1$ =1.549 Å<br>$b_2$ =0.774 Å<br>$E$ = -3.609 eV |
| $B_6$ | $C_{5v}$ | 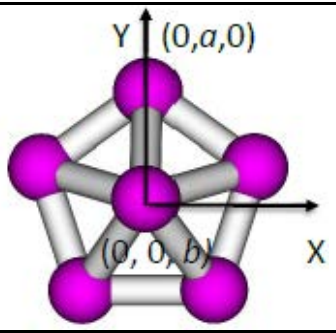 | $a$ =1.345 Å<br>$b$=0.943 Å<br>$E$ = -3.341 eV | $a$ =1.365 Å<br>$b$ =0.932 Å<br>$E$ = -3.741 eV |
| $B_6$ | $C_{2h}$ | 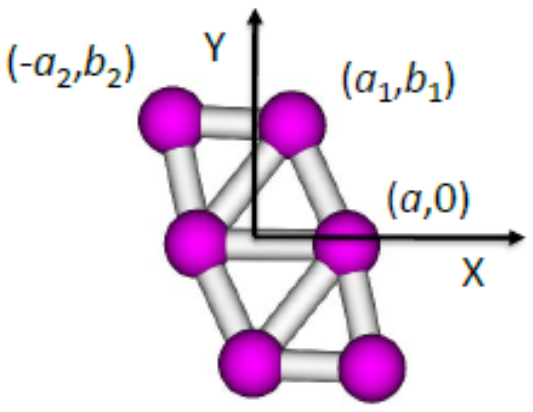 | $a$ =0.853 Å<br>$a_1$ =0.153 Å<br>$a_2$ =1.237 Å<br>$b_1$=1.433 Å<br>$b_2$ =1.491 Å<br>$E$ = -3.273 eV | $a$ =0.916 Å<br>$a_1$=0.220 Å<br>$a_2$=1.252 Å<br>$b_1$=1.452 Å<br>$b_2$ =1.501 Å<br>$E$ = -3.585 eV |
| $B_7$ | $C_{2v}$ | 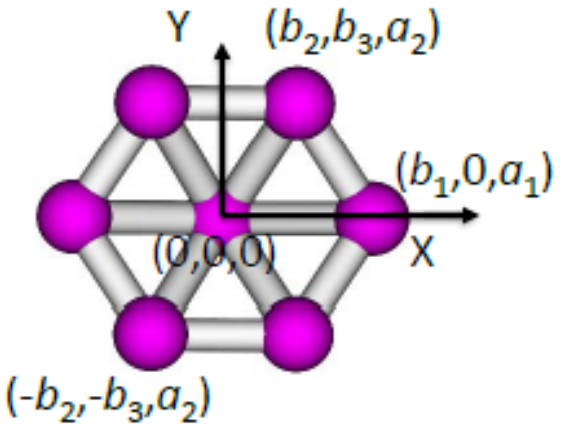 | $a_1$ =0.498 Å<br>$a_2$=0.576 Å<br>$b_1$ =1.607 Å<br>$b_2$ =0.812 Å<br>$b_3$ =1.340 Å<br>$E$ = -3.560 eV | $a_1$ =0.494 Å<br>$a_2$ =0.708 Å<br>$b_1$=1.654 Å<br>$b_2$ =0.809 Å<br>$b_3$ =1.300 Å<br>$E$ = -3.463 eV |
| $B_7$ | $C_{2h}$ | 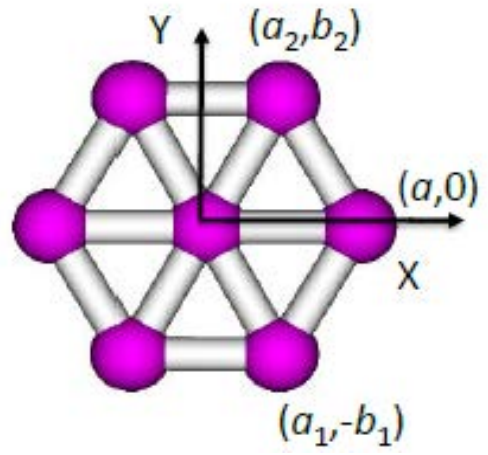 | $a$=1.626 Å<br>$a_1$ =1.433 Å<br>$a_2$=1.385 Å<br>$b_1$=0.770 Å<br>$b_2$ =0.852 Å<br>$E$ = -3.537 eV | $a$=1.621 Å<br>$a_1$ =1.427 Å<br>$a_2$ =1.420 Å<br>$b_1$ =0.781 Å<br>$b_2$ =0.797 Å<br>$E$ = -3.405 eV |
| $B_8$ | $D_{8h}$ | 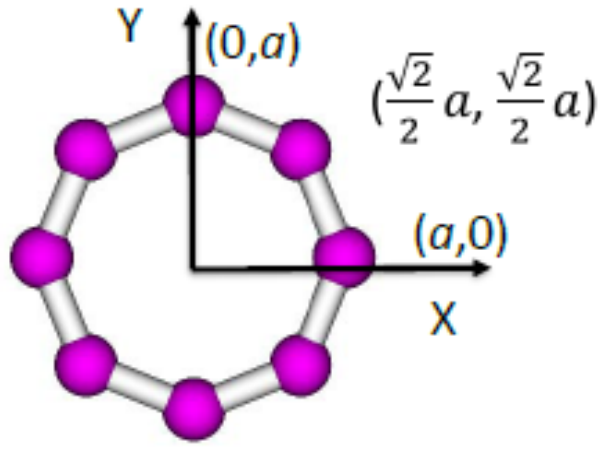 | $a$ =2.031 Å<br>$E$ = -3.158 eV | $a$ =2.012 Å<br>$E$ = -3.043 eV |
| $B_9$ | $D_{7h}$ | 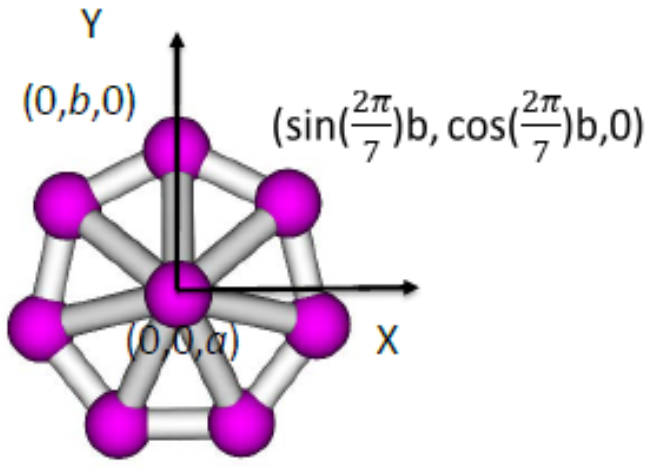 | $a$ =0.786 Å<br>$b$ =1.686 Å<br>$E$ = -3.781 eV | $a$ =0.853 Å<br>$b$ =1.755 Å<br>$E$ = -3.721 eV |
| $B_{12}$ | $C_{3v}$ | 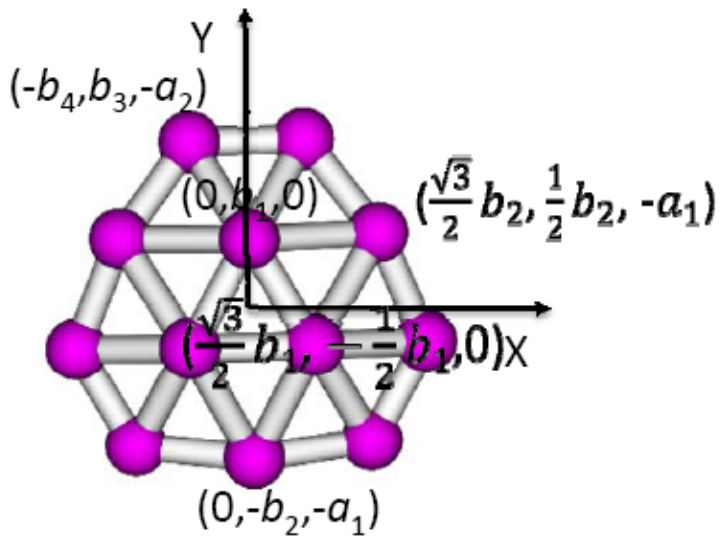 | $a_1$=0.429 Å<br>$a_2$=0.552 Å<br>$b_1$ =0.955 Å<br>$b_2$=1.979 Å<br>$b_3$ =2.284 Å<br>$b_4$=0.773 Å<br>$E$ = -3.985 eV | $a_1$ =0.430 Å<br>$a_2$=0.559 Å<br>$b_1$ =0.968 Å<br>$b_2$ =2.023 Å<br>$b_3$=2.290 Å<br>$b_4$ =0.774 Å<br>$E$ = -4.020 eV |
| $B_{12}$ | $D_{6d}$ | 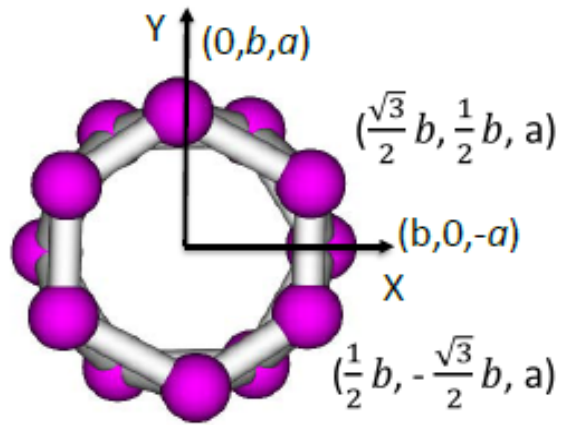 | $a$ =0.726 Å<br>$b$ =1.613 Å<br>$E$ = -3.890 eV | $a$ =0.731 Å<br>$b$ =1.618 Å<br>$E$ = -3.855 eV |

Table 3 The relative cohesive energy (eV) values per atom for boron clusters $B_N$ ( $N$ =10, 11, 12, 13, 15, 19, 20, 24, 36, and 40) and their isomers as calculated from the SCED-LCAO method and the DFT-GGA method, as implemented in the VASP [45] are reported. The cohesive energy of the B40-2 isomer was used as the reference value in calculating the relative cohesive energies.

| $B_N$ clusters | Structures | SCED-LCAO results | DFT results [17] |
|---|---|---|---|
| $B_{10}$ | 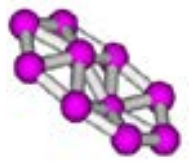 | 0.38885 | 0.53775 |
| $B_{11}$ | 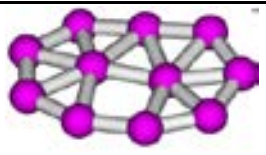 | 0.37145 | 0.48303 |
| B12 | 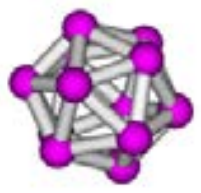 | 0.38978 | 0.67599 |
| $B_{13\text{-}1}$ | 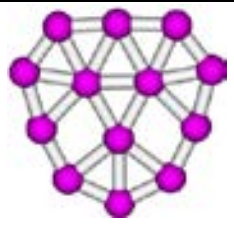 | 0.32157 | 0.40298 |
| $B_{13\text{-}2}$ | 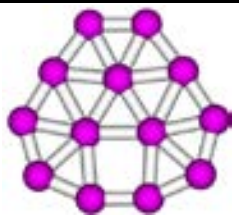 | 0.32163 | 0.40323 |
| $B_{15\text{-}1}$ | 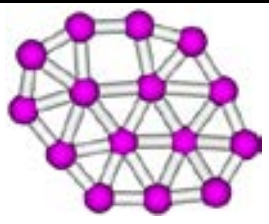 | 0.27756 | 0.35082 |
| $B_{15\text{-}2}$ | 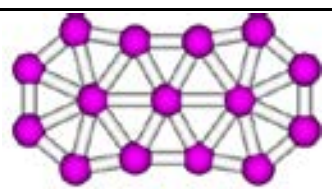 | 0.26723 | 0.34406 |
| $B_{19}$ | 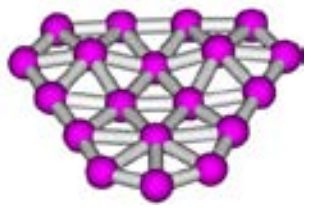 | 0.17515 | 0.23592 |
| $B_{20}$ | 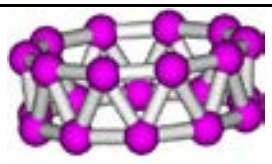 | 0.12021 | 0.18017 |
| $B_{24}$ | 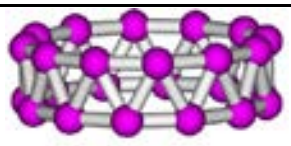 | 0.07391 | 0.11637 |
| $B_{36\text{-}1}$ | 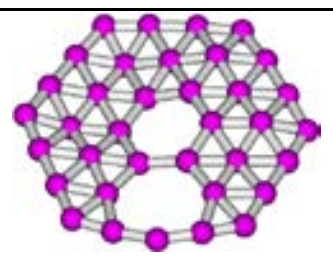 | 0.03945 | 0.03248 |
| $B_{36\text{-}2}$ | 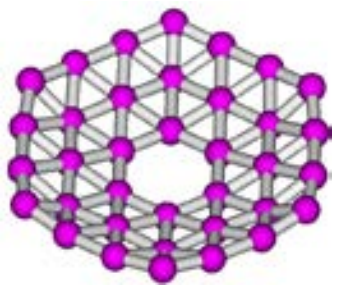 | 0.00839 | 0.00098 |
| $B_{40\text{-}1}$ | 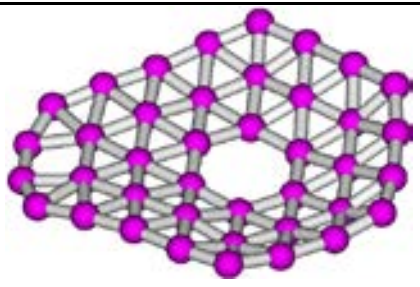 | 0.01420 | 0.00598 |
| $B_{40\text{-}2}$ | 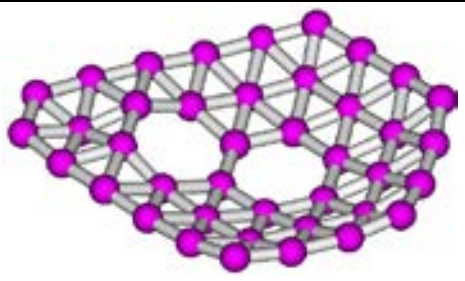 | 0.0 | 0.0 |

Table 4 The relative cohesive energies per atom for different types of two-dimensional boron sheets are reported with respect to the cohesive energy of the bulk alpha boron for the SCED-LCAO and DFT-GGA (values in the parenthesis [45]). The calculated energy gap values of these structures using SCED-LCAO and DFT-GGA (parenthesis [45]) are also given.

| Symmetry | Structure | Relative cohesive energy (eV/atom) | Energy gap (eV) |
|---|---|---|---|
| Flat $\delta_4$ sheet | 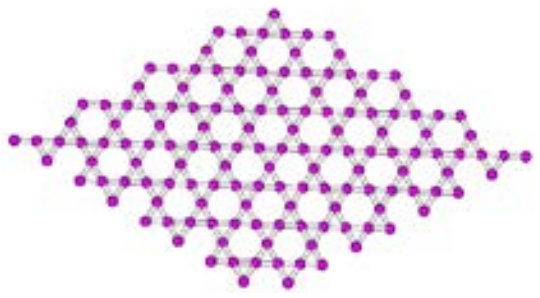 | 0.613 (0.891) | Gapless (gapless) |
| Flat $B_{36}$ sheet (borophene) | 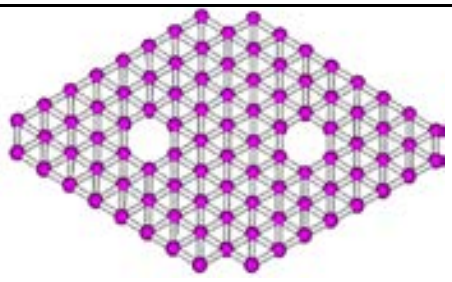 | 0.476 (0.605) | 0.131 (0.122) |
| Buckled triangle sheet | 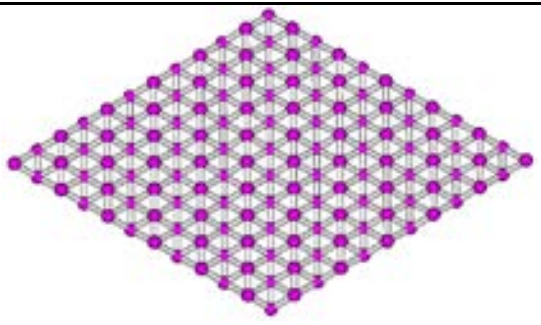 | 0.421 (0.496) | Gapless (gapless) |
| Buckled α sheet | 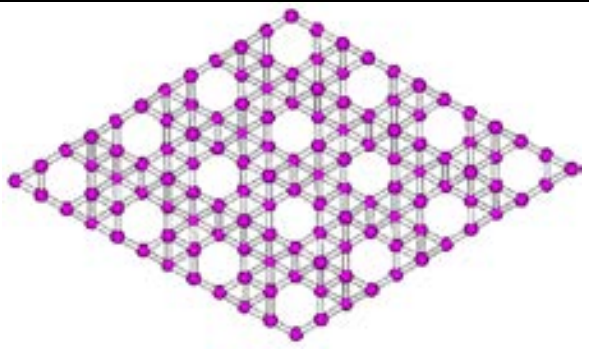 | 0.356 (0.378) | Gapless (gapless) |
| alpha boron | 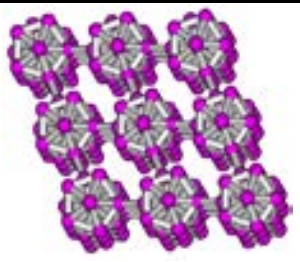 | 0.0 (0.0) | 1.902 (1.953) |

Table 5 Optimized structures of intermediate-sized boron clusters of similar sizes generated from different initial configurations and bulk cuts. The relative energy per atom of these structures and their corresponding geometric properties, namely, the average bond length (Å) and the average number of neighbors in the cluster using a cut-off radius of 2.0 Å are reported.

| structure | Average bond length | Average neighbors | relative energy |
|---|---|---|---|
| $B_{216}$ (9_2_1) 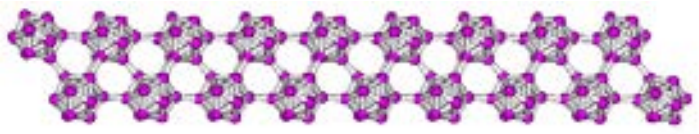 | 1.697 | 5.379 | 0.503 |
| $B_{228}$ (Random) 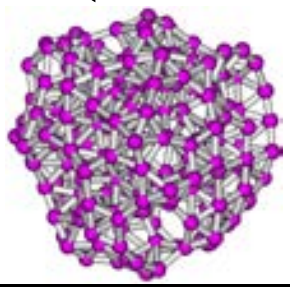 | 1.723 | 5.833 | 0.415 |
| $B_{210}$ (Empty) 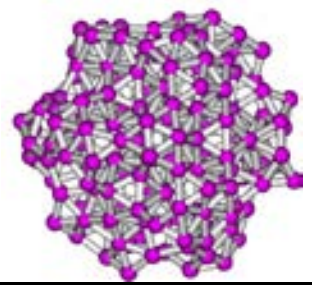 | 1.722 | 5.570 | 0.408 |
| $B_{216}$ (Ball) 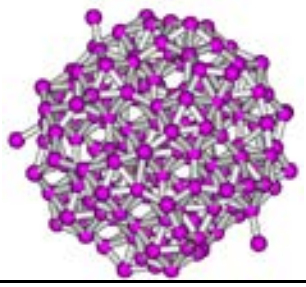 | 1.728 | 5.810 | 0.39 |
| $B_{216}$ (3_3_2) 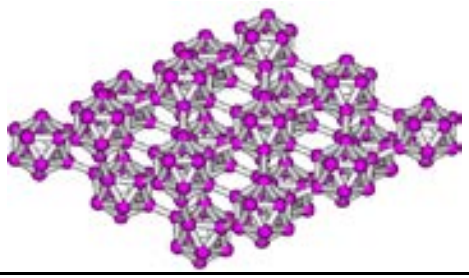 | 1.715 | 5.676 | 0.361 |
| $B_{228}$ (Trimmed) 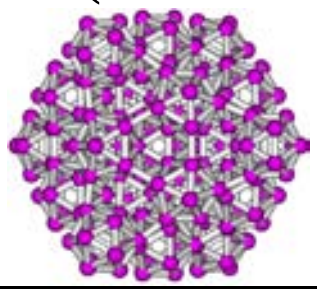 | 1.722 | 5.842 | 0.319 |
| alpha boron 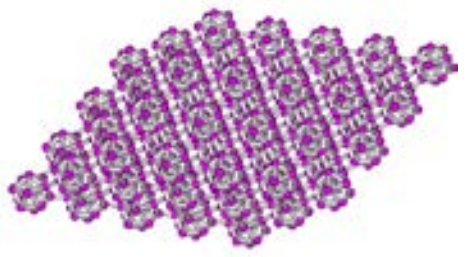 | 1.774 | 5.500 | 0.0 |

**Figure captions**

Figure 1 (Color on line) The relative cohesive energy per atom versus the ratio of the atomic volume ($V/V_0$) corresponding to different extended phases: (i) the bulk alpha boron, (ii) the α-sheet, and (iii) the $\delta_4$-sheet as calculated using the SCED-LCAO (solid curves) and the DFT-GGA [45] . The relative energy per atom $E - E_\alpha^0$ is defined as the difference between the total energy per atom of any extended phase (E) with respect to that of the bulk α-boron phase at its equilibrium volume $V_0$. The optimized structures corresponding to the bulk alpha-boron phase (left), the α-sheet (middle), and the $\delta_4$-sheet (right) are also shown.

Figure 2 (Color on line) The total energy per atom versus MD steps for $B_{80}$ buckyball at 0 K. The buckyball with the initial $I_h$ symmetry (top left) relaxes to the $T_h$ symmetry (middle) in 0.3 ps, and it stays in the $T_h$ symmetry for about 1.2 ps. It finally stabilizes to the $C_{2h}$ symmetry (bottom right) after 2.1 ps. The direction of the arrow is used to denote whether the central atom of the boron hexagonal rings is protruded inwards or outwards.

Figure 3 (Color on line) Annealing and cooling of the $B_{80}$ buckyball structure (0K-> 1800K->0K) renders the optimized $C_{2h}$ structure to transform into a random structure . The cohesive energy per atom of the random structure (-4.332 eV/atom) is 0.052 eV/atom lower compared to that of the $C_{2h}$ structure (-4.280 eV/atom).

Figure 4 (Color on line) The total energy per atom versus MD steps corresponding to the triangular sheet (black solid curve) and the α-sheet (red dashed curve), respectively. The initial flat triangular sheet (side and top views) stabilizes to a buckled triangular sheet (side view) in about 1.68 ps, while the initial flat α –sheet (side view) relaxes to a buckled α-sheet (side view) after 1.4 ps.

Figure 5 (Color on line) Optimized structures of three boron clusters exhibiting random configurations: $B_{101}$ (Random), $B_{228}$ (Random), and $B_{230}$ (Random).

Figure 6 (Color on line) The results for the optimized total energy per atom as a function of the cluster size (or number of atoms in the cluster) for several categories of initial configurations. (1) *chain-like clusters*: $B_N$ (i_1_1) (the black curve with open circles); (2) *rhombohedra sheet-like clusters*: $B_N$ (i_2_1) (the red curve with open squares) and $B_N$ (i_7_1) (the green curve with open diamonds); (3) *rhombohedra bulk-like clusters*: $B_N$ (i_2_2) (the red curve with solid squares), $B_N$ (i_3_2) (the green curve with solid diamonds), and $B_N$ (i_i_i) (the black curve with solid circles); (4) *spherical icosahedra clusters*: $B_N$ (Ball) (the blue open triangle-up), $B_N$ (Empty) ( the orange open triangle-down), and $B_{228}$ (Trimmed) (the black open square), and (5) *random $B_N$ (Random)*: the pink open circles. The result for the $B_{80}$ buckyball structure (black star) is also depicted. The initial structures of the rhombohedra clusters corresponding to $B_N$ (i_1_1) (top), $B_N$ (i_j_1) (middle), and $B_N$ (i_j_k) (bottom) are also illustrated.

Figure 7 (Color on line) Relaxed structures of several types of $B_N$ clusters with their initial structures derived from spherical truncations of the bulk alpha-boron of different radii with their centers located either at icosahedra sites (Notation: $B_N$ (Ball)) or at empty sites between icosahedra (Notation: $B_N$ (Empty)). The clusters on the left panel correspond to relaxed structures of $B_N$ (Ball) while on the right to $B_N$ (Empty).

Figure 8 (Color on line) The top and side views of the optimized structures of $B_{228}$ (trimmed) clusters (right panel) along with their corresponding initial structures (left panel). The notation $B_{228}$ (trimmed) refers to a cluster that is obtained from a spherical cut of the bulk alpha-boron with their dangling bonds trimmed.

Figure 9 (Color on line) The pair-distribution functions for $B_{216}$ (9_2_1) (the black short-dashed curve), $B_{228}$ (Random) (the red long-dashed curve), $B_{210}$ (Empty) (the green short-dot-dashed curve), $B_{216}$ (Ball) (the blue long-dot-dashed curve), $B_{216}$ (3_3_2) (the pink double dot-dashed curve), $B_{228}$ (Trimmed) (the black solid curve), and the alpha boron (the black dotted curve), respectively.

Figure 10 (Color on line) Angular distribution functions for $B_{216}$ (9_2_1) (the black short-dashed curve), $B_{228}$ (Random) (the red long-dashed curve), $B_{210}$ (Empty) (the green short-dot-dashed curve), $B_{216}$ (Ball) (the blue long-dot-dashed curve), $B_{216}$ (3_3_2) (the pink double dot-dashed curve), $B_{228}$ (Trimmed) (the black solid curve), and the alpha boron (the black dotted curve), respectively.

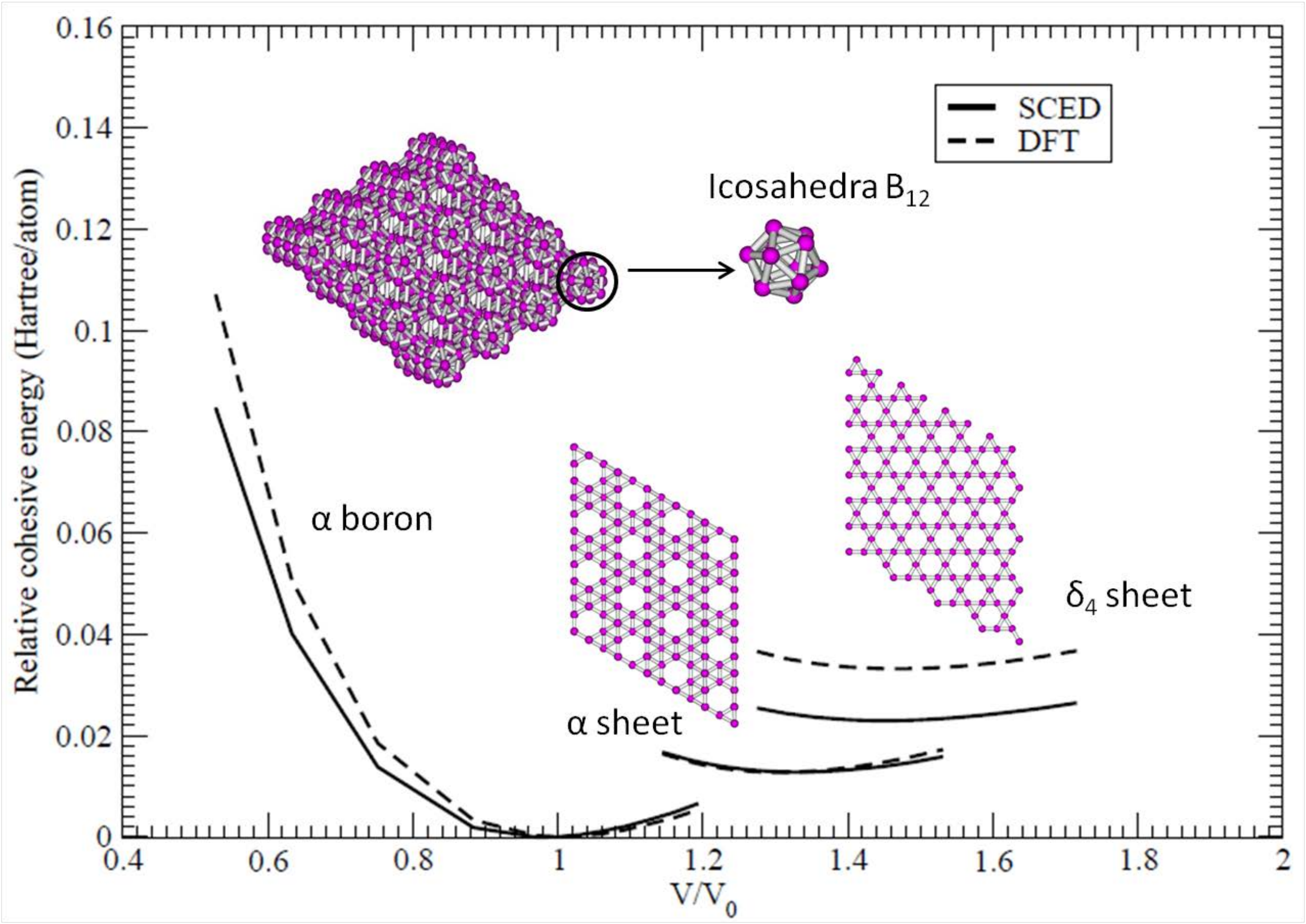

0.16
0.14
0.12
0.1
0.08
0.06
0.04
0.02
0
0.4
0.6
0.8
1
1.2
1.4
1.6
1.8
2
Relative cohesive energy (Hartree/atom)
V/V0
SCED
DFT
Icosahedra B12
α boron
α sheet
δ4 sheet

Figure 1

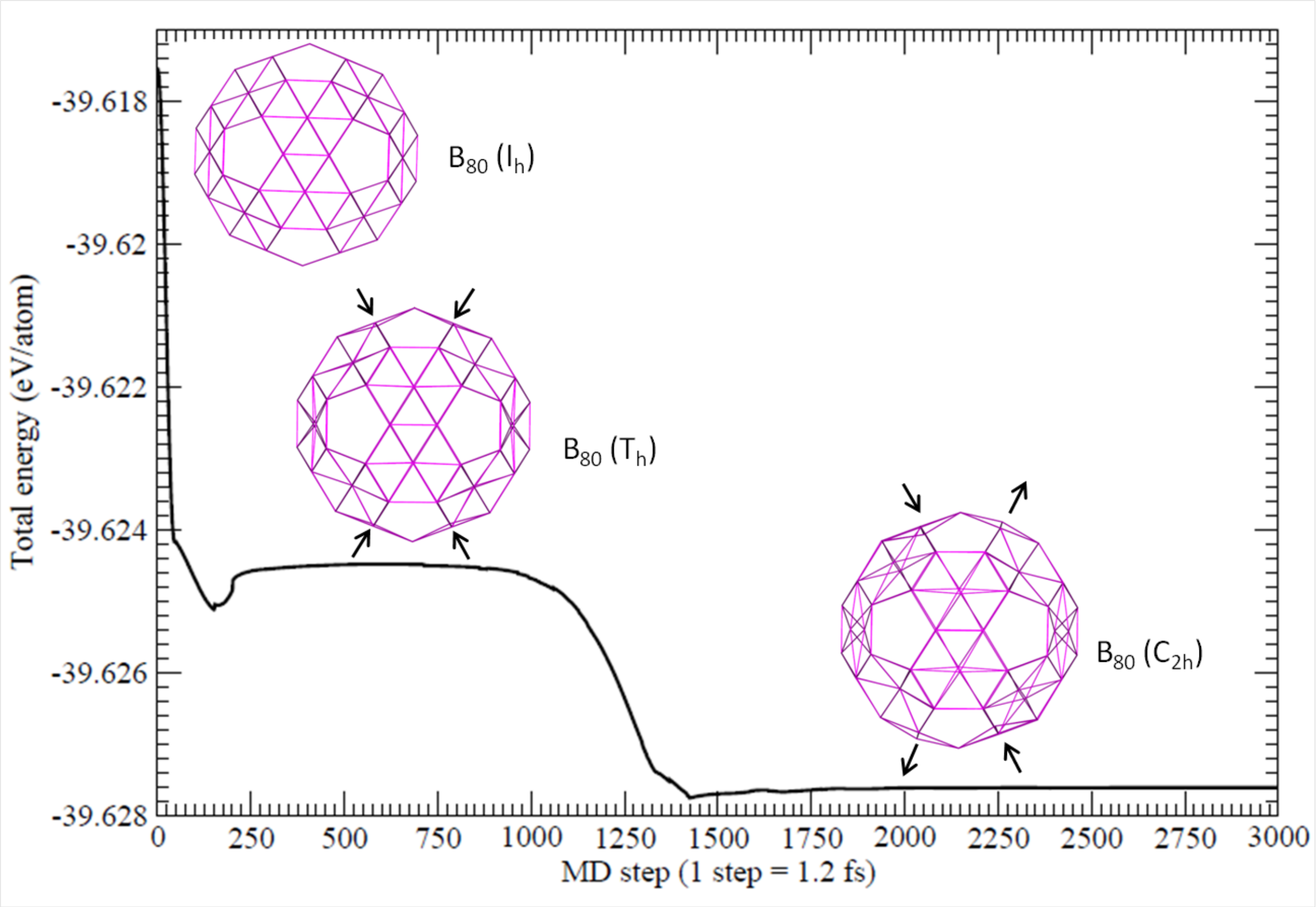

Figure 2

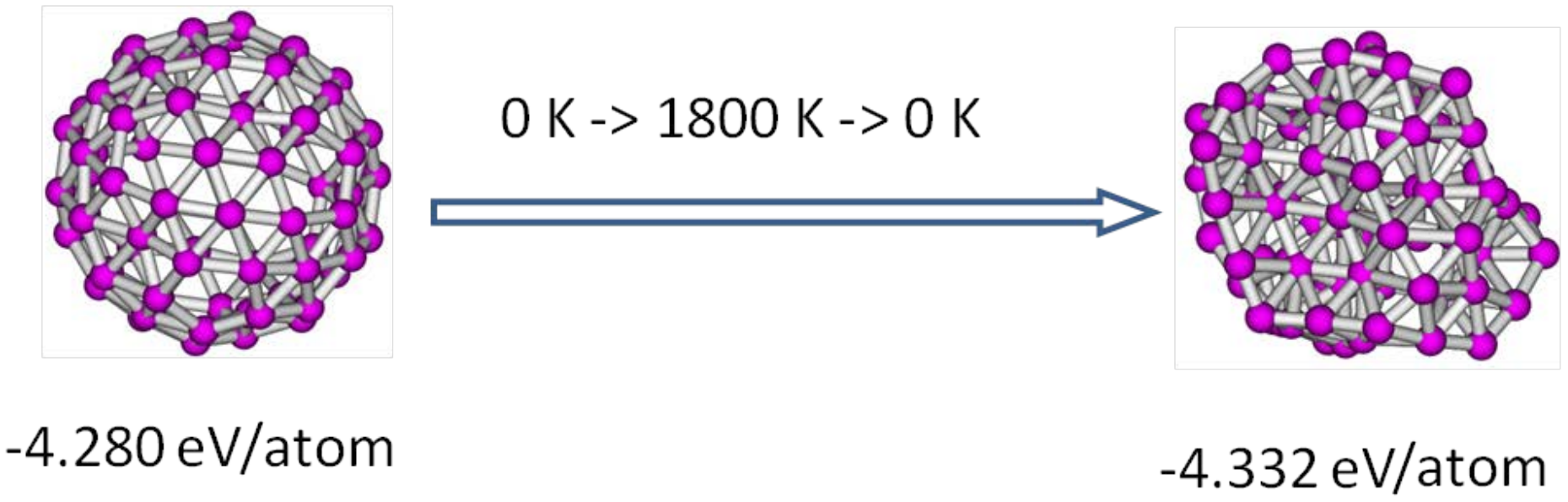

Figure 3

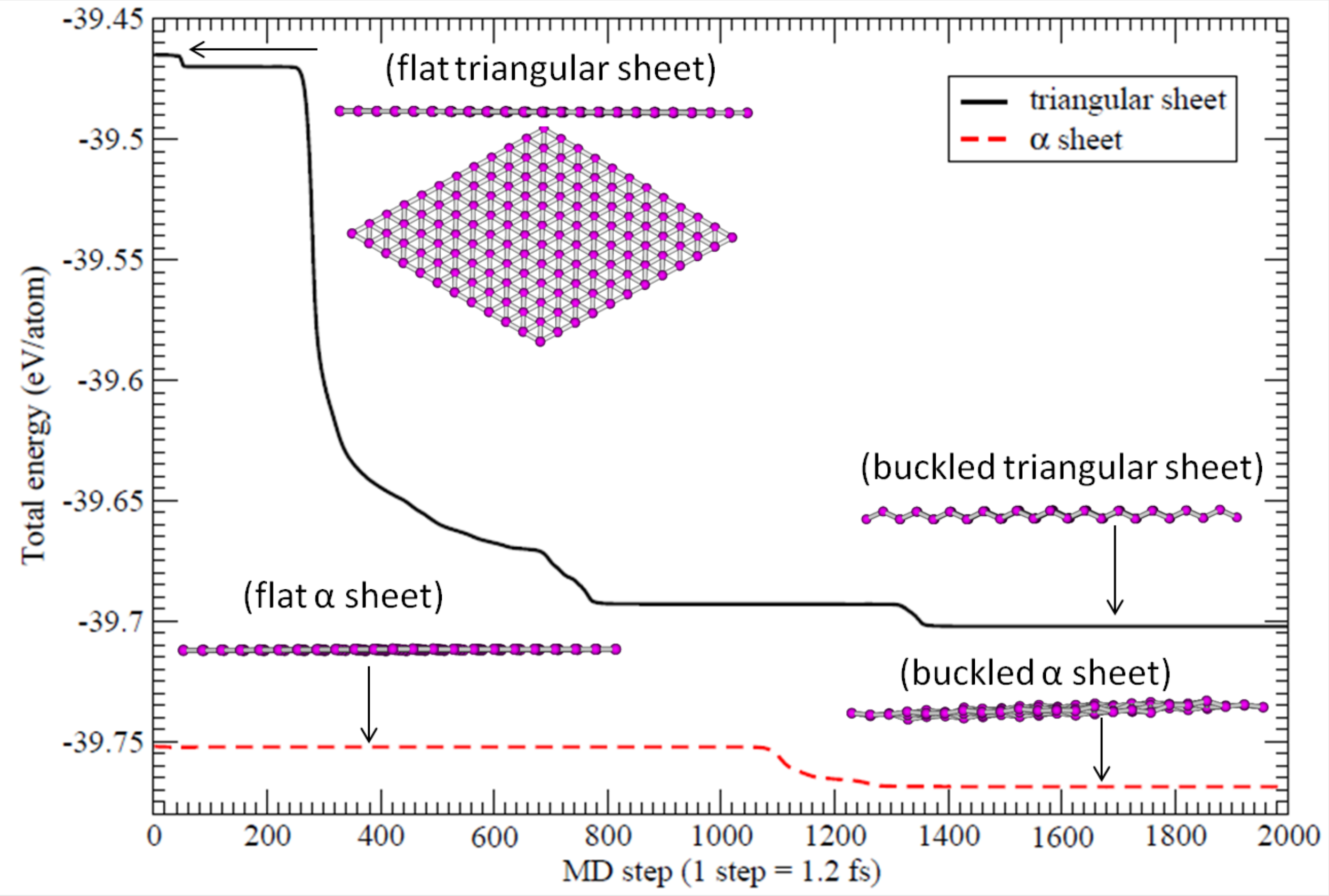
(flat triangular sheet)
triangular sheet
α sheet
(buckled triangular sheet)
(flat α sheet)
(buckled α sheet)
Total energy (eV/atom)
-39.45
-39.5
-39.55
-39.6
-39.65
-39.7
-39.75
0
200
400
600
800
1000
1200
1400
1600
1800
2000
MD step (1 step = 1.2 fs)

Figure 4

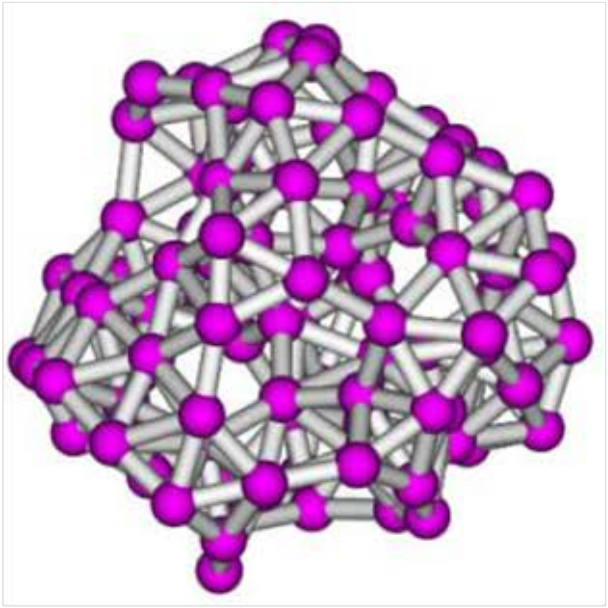

$B_{101}$ (Random)

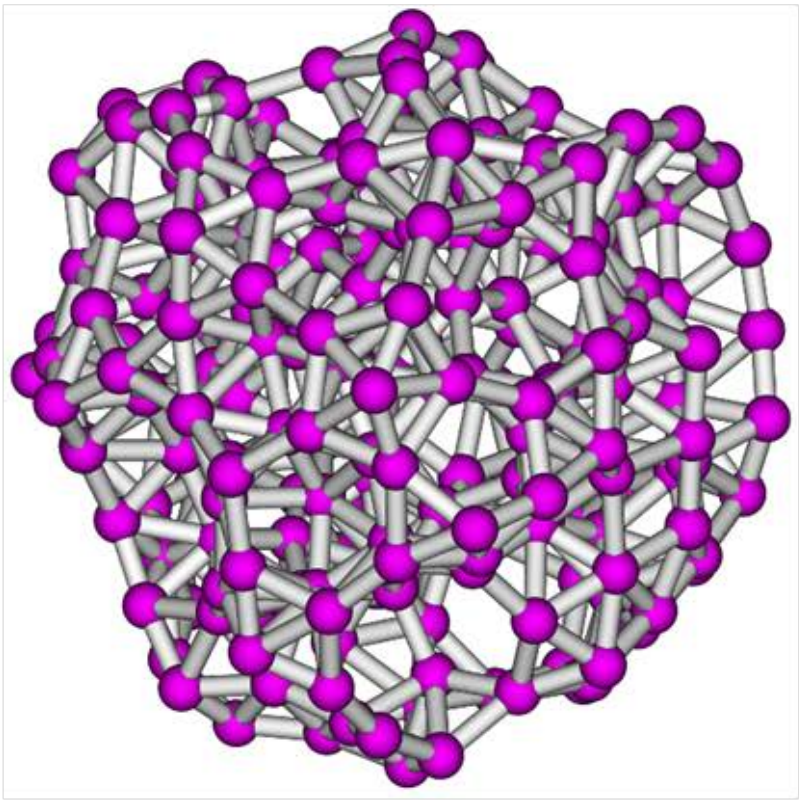

$B_{228}$(Random)

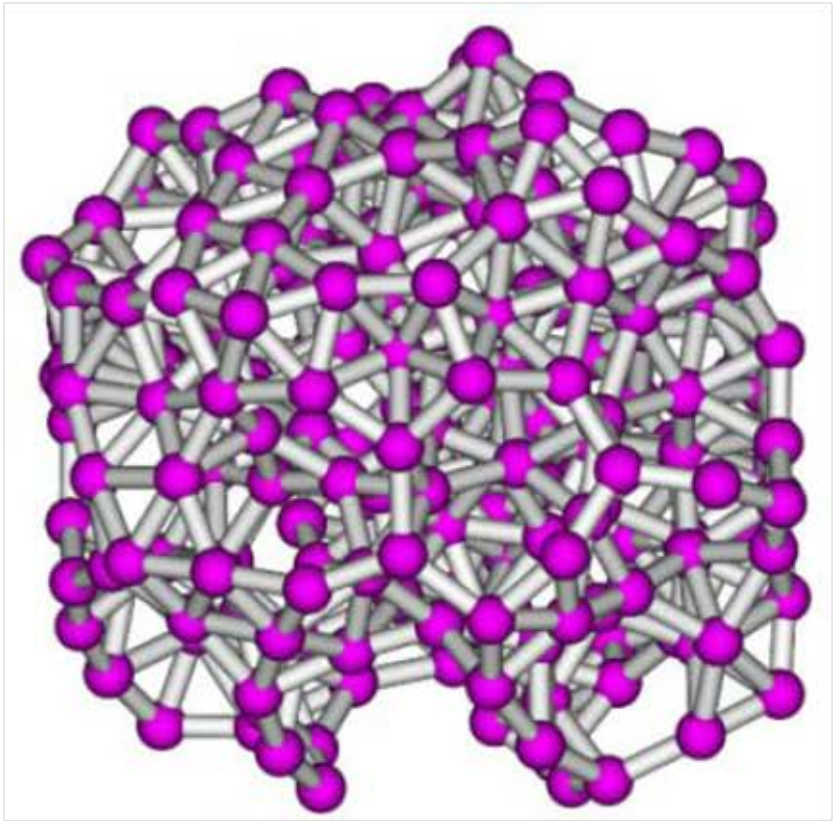

$B_{230}$(Random)

Figure 5

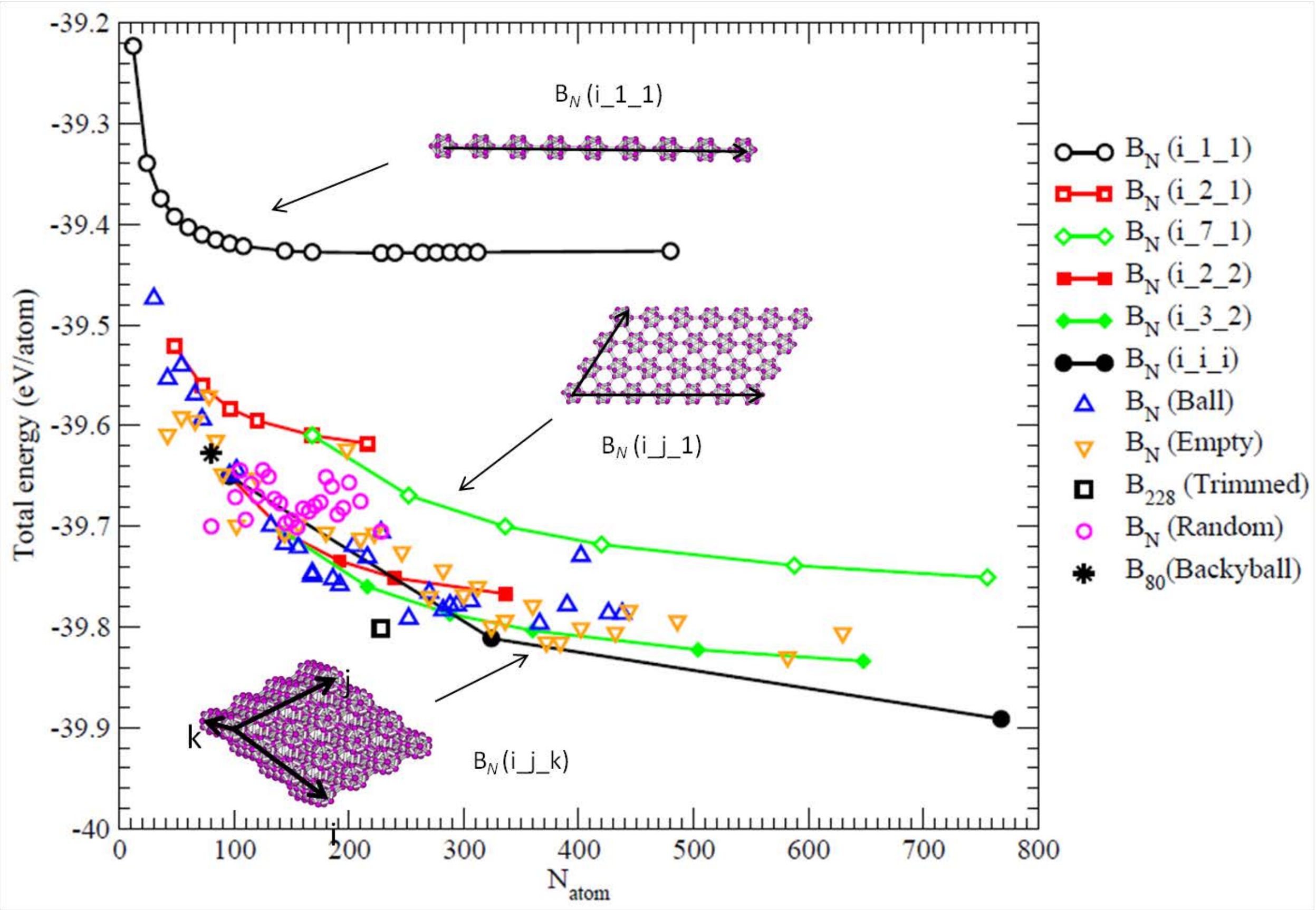

-39.2
-39.3
-39.4
-39.5
-39.6
-39.7
-39.8
-39.9
-40
Total energy (eV/atom)
0
100
200
300
400
500
600
700
800
N_atom
B_N (i_1_1)
B_N (i_j_1)
B_N (i_j_k)
j
k
i
B_N (i_1_1)
B_N (i_2_1)
B_N (i_7_1)
B_N (i_2_2)
B_N (i_3_2)
B_N (i_i_i)
B_N (Ball)
B_N (Empty)
B_228 (Trimmed)
B_N (Random)
B_80(Backyball)

Figure 6

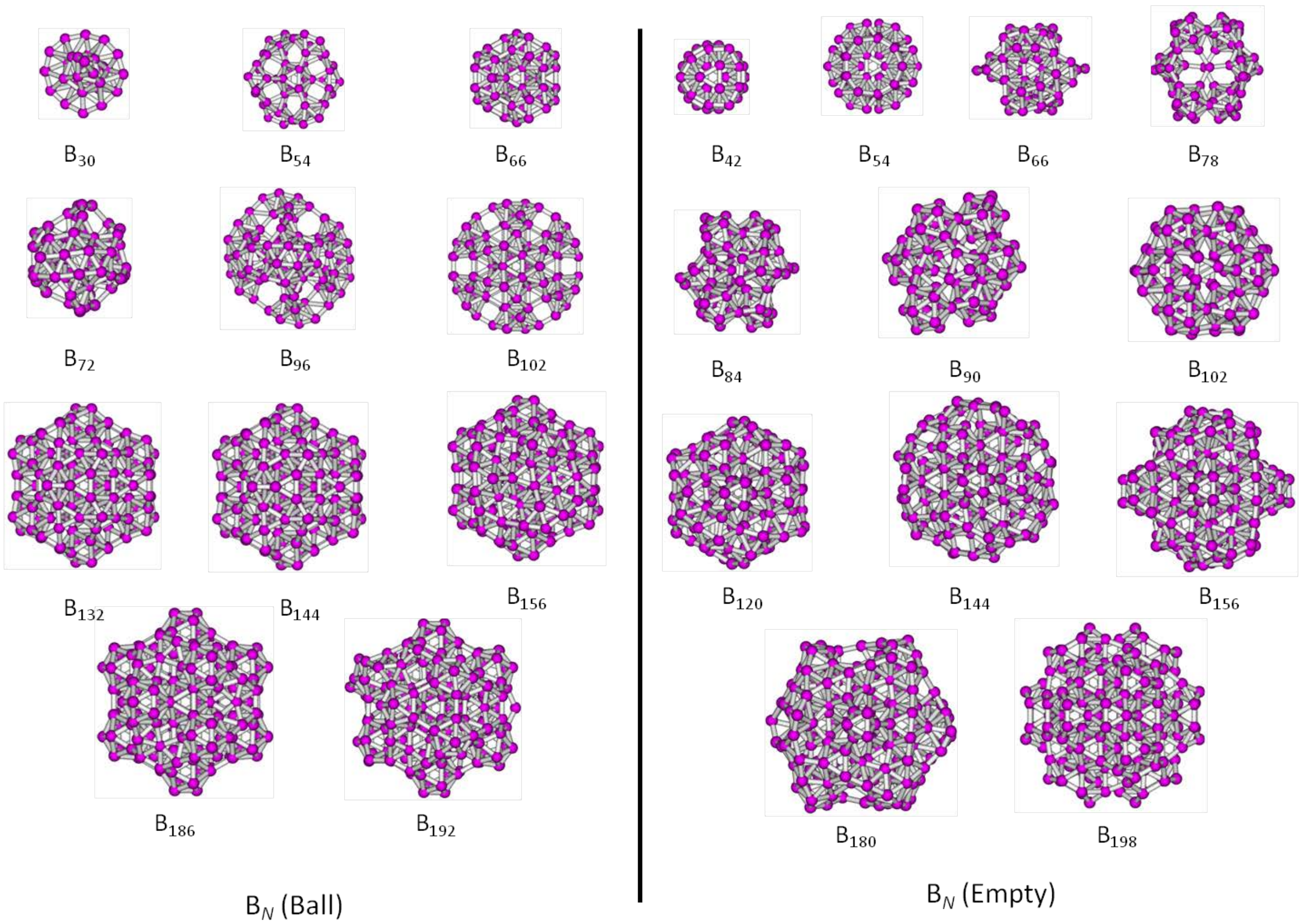
$B_{30}$
$B_{54}$
$B_{66}$
$B_{72}$
$B_{96}$
$B_{102}$
$B_{132}$
$B_{144}$
$B_{156}$
$B_{186}$
$B_{192}$
$B_N$ (Ball)
$B_{42}$
$B_{54}$
$B_{66}$
$B_{78}$
$B_{84}$
$B_{90}$
$B_{102}$
$B_{120}$
$B_{144}$
$B_{156}$
$B_{180}$
$B_{198}$
$B_N$ (Empty)

Figure 7

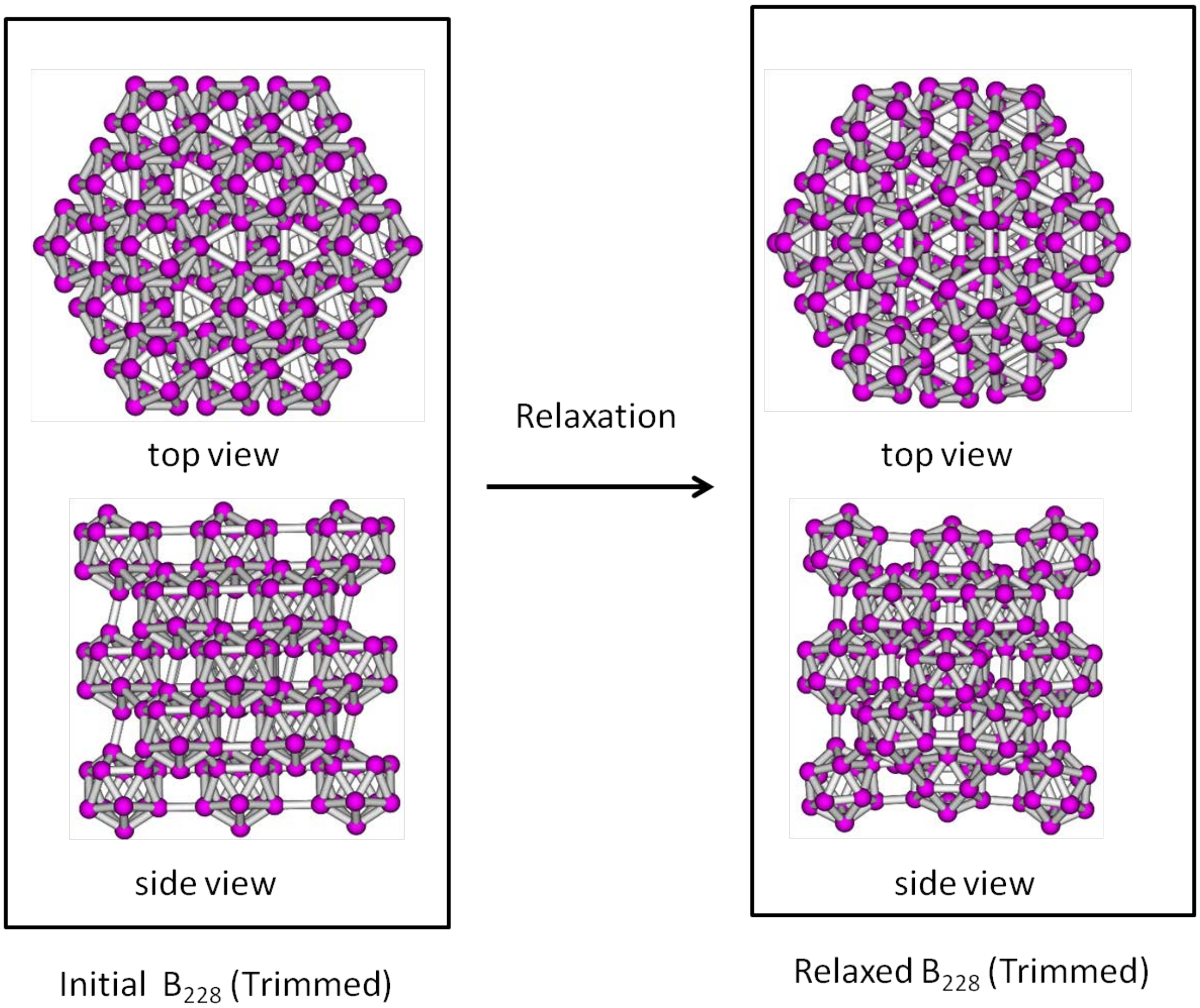
top view
side view
Initial $B_{228}$ (Trimmed)
Relaxation
top view
side view
Relaxed $B_{228}$ (Trimmed)

Figure 8

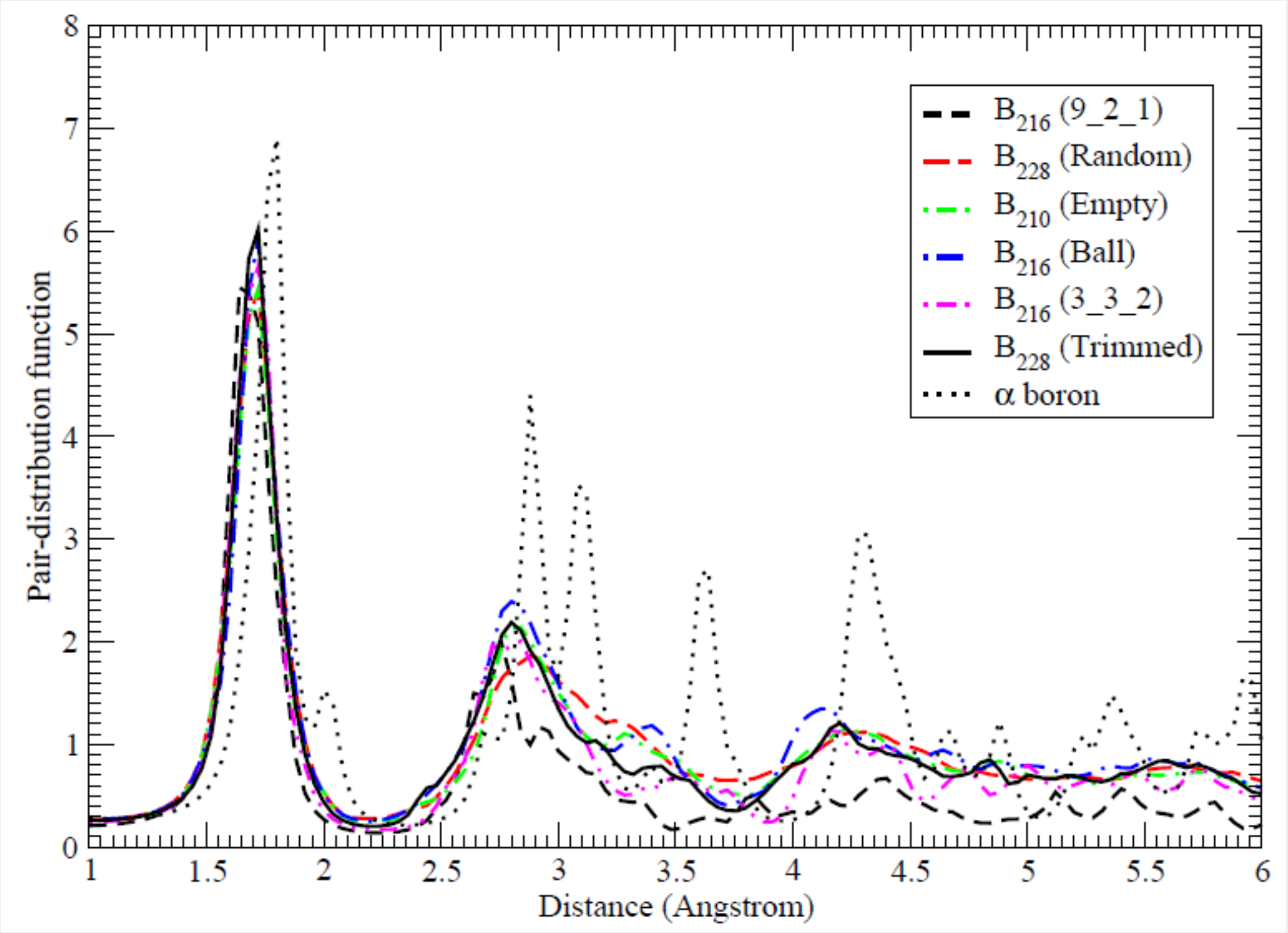

Figure 9

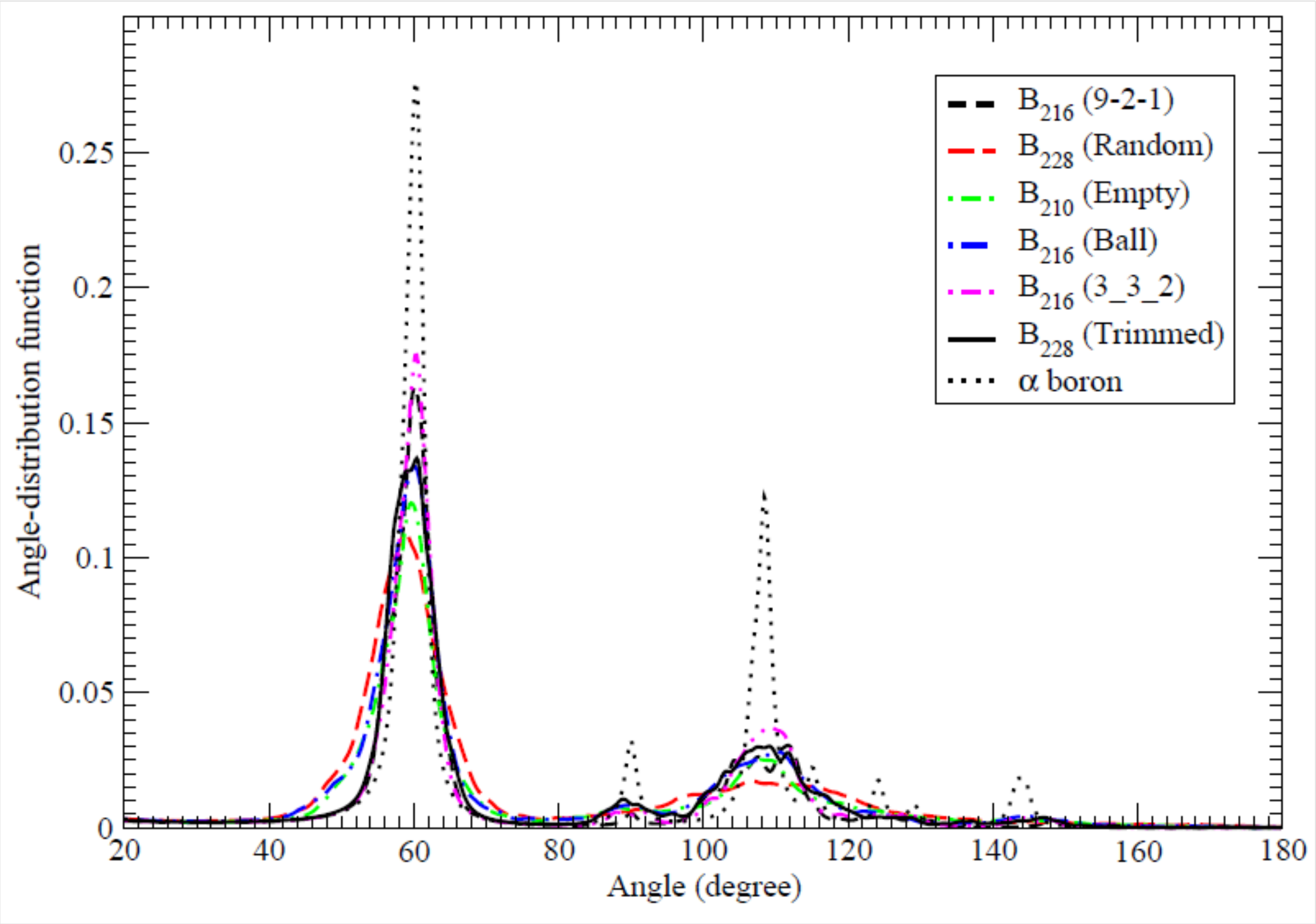

Figure 10